\documentclass{aa}  
\usepackage{natbib}
\bibpunct{(}{)}{;}{a}{}{,} 
\usepackage{txfonts}
\usepackage{tasks}
\usepackage{float}
\usepackage{graphicx}
\usepackage{booktabs}
\usepackage{dblfloatfix}
\usepackage{hyperref}
\usepackage{afterpage}
\usepackage{amsmath}
\graphicspath{ {images/} }

\begin{document} 

   \title{
   Radio emission in a nearby ultracool dwarf binary: a multi-frequency study
   }


   \author{J. B. Climent
          \inst{1}\fnmsep\inst{2}\thanks{Email: j.bautista.climent@uv.es}
          \and
          J. C. Guirado\inst{1}\fnmsep\inst{3}
          \and
          M. R. Zapatero Osorio\inst{4}
          \and
          O. V. Zakhozhay\inst{5}
          \and
          M. P\'erez-Torres\inst{6}
          \and
          R.~Azulay\inst{3}          
          \and 
          B.~Gauza\inst{7}\fnmsep\inst{8}
          \and
          R. Rebolo\inst{9}\fnmsep\inst{10}\fnmsep\inst{11}
          \and
          V. J. S. B\'ejar\inst{9}\fnmsep\inst{10}
          \and
          J.~Mart\'in-Pintado\inst{4}
          \and 
          C. Lef\`evre\inst{12}
          }

         \institute{Departament d’Astronomia i Astrof\'isica, Universitat de Val\`encia, C. Dr. Moliner 50, E-46100 Burjassot, Val\`encia, Spain.
         \and
         Universidad Internacional de Valencia (VIU), C/ Pintor Sorolla 21, E-46002 Valencia, Spain
        \and
         Observatori Astron\`omic, Universitat de Val\`encia, Parc Cient\'ific, C. Catedr\'atico Jos\'e Beltr\'an 2, E-46980 Paterna, Val\`encia, Spain.
         \and
         Centro de Astrobiolog\'ia (CSIC-INTA), Carretera de Ajalvir km 4, 28850 Torrej\'on de Ardoz, Madrid, Spain.
         \and 
         Main Astronomical Observatory, National Academy of Sciences of Ukraine, Kyiv 03680, Ukraine.
         \and
         Instituto de Astrof\'isica de Andaluc\'ia, Consejo Superior de Investigaciones Cient\'ificas (CSIC), Glorieta de la Astronom\'ia s/n, E-18008, Granada, Spain.
         \and 
         Centre for Astrophysics Research, University of Hertfordshire, College Lane, Hatfield AL10 9AB, UK.
         \and
         Janusz Gil Institute of Astronomy, University of Zielona G\'ora, Lubuska 2, 65-265 Zielona G\'ora, Poland.
         \and 
         Instituto de Astrof\'isica de Canarias, E-38205, La Laguna, Tenerife, Spain.
         \and 
         Universidad de La Laguna, Departamento de Astrof\'isica, La Laguna, Tenerife, E-38206, Spain.
         \and
         Consejo Superior de Investigaciones Cient\'ificas, CSIC, Spain.
         \and
         Institut de Radioastronomie Millim\'etrique (IRAM), 38406 Saint-Martin-d’H\'eres, France
         }

   \date{Received ...; accepted ...}

 
  \abstract
   {
   The substellar triple system VHS J125601.92$-$125723.9 (hereafter VHS 1256$-$1257) is composed by an equal-mass M7.5 brown dwarf binary and a L7 low-mass substellar object. In \citet{2018Guirado} we published the detection of radio emission at 8.4 GHz coming from the central binary and making it an excellent target for further observations. 
   }
   {
   We aim to identify the origin of the radio emission occurring in the central binary of VHS 1256$-$1257 while discussing the expected mechanisms involved in the radio emission of ultracool dwarfs (UCDs).
   }
   {
   We observed this system with the \textit{Karl G. Jansky} Very Large Array, the European very-long-baseline interferometry (VLBI) Network, the enhanced Multi Element Remotely Linked Interferometer Network, the NOrthern Extended Millimeter Array, and the Atacama Large Millimetre Array at frequencies ranging from 5 GHz up to 345 GHz in several epochs during 2017, 2018, and 2019.
   }
   {
   We have found radio emission at 6 GHz and 33 GHz coincident with the expected position of the central binary of VHS~1256$-$1257. The Stokes I density flux detected were 73 $\pm$ 4 $\mu$Jy and 83 $\pm$ 13 $\mu$Jy, respectively, with no detectable circular polarisation or pulses. 
   No emission is detected at higher frequencies (230 GHz and 345 GHz) nor at 5 GHz with VLBI arrays.
   The emission appears to be stable over almost 3 years at 6 GHz.
   To explain the constraints obtained both from the detections and non-detections we considered multiple scenarios including thermal and non-thermal emission, and different contributions from each component of the binary.
   } 
   {
   Our results can be well explained by non-thermal gyrosynchrotron emission originating at radiation belts with a low plasma density (n$_e$ = 300$-$700 cm$^{-3}$), a moderate magnetic field strength  (B $\approx$ 140 G), and an energy distribution of electrons following a power-law ($d N / d E \propto E^{- \delta}$) with $\delta$ fixed at 1.36. These radiation belts would need to be present in both components and also be viewed equatorially. 
   }


   \maketitle
%
\section{Introduction}\label{sect:introduction}

Ultracool dwarfs (UCDs) are stellar and substellar objects with spectral type later than M7 \citep{1997Kirkpatrick}. Due to their low masses and temperatures, UCDs were thought to lack the Sun-like dynamo and, consequently, any strong magnetic field \citep{2002Mohanty}. However, this consensus was abandoned since the discovery of radio emission from the M9 object LP~944-20 \citep{2001Berger}. Surveys of UCDs have shown that up to $\sim$10\% exhibit radio emission \citep{2016Route} whose origin is attributed to a combination of gyrosynchrotron radiation \citep[explaining the quiescent emission;][]{2002Berger} and the electron cyclotron maser instability \citep[ECMI; explaining the detected highly-polarized pulses;][]{2007Hallinan,2008Hallinan}. Interestingly, quiescent emission always accompanies pulse radio emission \citep[e.g.][]{2009Berger,2016Kao} but not vice versa \citep[e.g.][]{2006Berger}.

Despite their lack of $\alpha\Omega$ dynamo, UCDs have been confirmed to possess surface-averaged magnetic field strength of order kilogauss via Zeeman broadening and Zeeman Doppler imaging \citep[e.g.][]{2006Donati,2010Reinersbasri,2017Shulyak}, in agreement with estimations from observed pulsed radio emission \citep[e.g.][]{2016Route,2016Kao} and gyrosynchrotron emission \citep[e.g.][]{2018Guirado}. The underlying mechanisms responsible for such strong magnetic fields are still unknown but a handful of models have been proposed in the literature \citep[e.g.][]{2008Browning,2009Christensen,2009Simitev,2011Morin,2013Gastine}.

Observations of UCDs at radio-wavelengths are a powerful tool for probing the magnetic activity of these objects and, in the case of late L- and T-type objects, it is the only valid tool we currently possess. Additionally, the knowledge gathered from such observations may open a suitable route to the detection of exoplanetary radio emission. 
However, in-depth studies of UCD radio emission are still relatively scarce due, in part, to their difficult detection. New observations are needed to distinguish among the different proposed mechanisms for the origin of the strong magnetic fields and radio emission detected in these objects.
As such, the system  VHS J125601.92–125723.9 \citep[hereafter  VHS 1256–1257;][]{2015Gauza} represents an excellent opportunity since its radio emission has been previously confirmed \citep{2018Guirado} and new observations can provide further constraints on its origin. 
This system is relatively nearby with the most recent measured parallaxes being 45.0 $\pm$ 2.4 mas \citep[][]{2020Dupuy} and 47.3 $\pm$ 0.5 mas \citep{2021Gaia}.
It is composed by a 0.1$"$ equal-magnitude  M7.5  binary \citep[VHS 1256$-$1257A and VHS 1256$-$1257B;][]{2016Stone} and  a  lower  mass L7  companion  \citep[component  b;][]{2016Rich} located 8$"$ away from the central pair. 
It is one of the few systems in which all three components are substellar \citep{2005Bouy,2013Radigan}. The masses of the central pair components are estimated to be 50-90 M$_{\mathrm{Jup}}$ each and 10-35 M$_{\mathrm{Jup}}$ for the L7 companion \citep{2015Gauza,2016Rich,2016Stone,2018Guirado, 2020Dupuy}. This locates VHS 1256$-$1257b on the planet-brown dwarf boundary.
The spectroscopic and photometric characteristics of VHS~1256$-$1257b resemble those of the free-floating planetary-mass objects WISEJ0047 \citep{2012Gizis,2016Lew} and PSOJ318 \citep{2013Liu,2018Biller}, and the exoplanets HR8799bcde \citep{2008Marois,2010Marois}. 

The strong  lithium  depletion  observed  in  the  high  resolution  spectra  of  the central pair and its kinematic membership to the Local Association implied an age of 150–300 Myr \citep{2015Gauza,2016Rich,2016Stone}. This young age, together with the 102 AU separation between the central pair and the L7 object, makes VHS 1256$-$1257 one of the most suitable system to search for debris disc around UCDs and exoplanets. Sub-mm observations could probe not only the emission of cold dust surrounding the central binary but also  detect a dusty disc surrounding an L-type object, as suggested for others like G196$-$3B \citep{2017Zakhozhay}.

Previous radio observations of VHS 1256$-$1257 have shown emission coincident with the central binary at 8.4 GHz \citep[peak density of $\sim$60 $\mathrm{\mu Jy}$/beam;][]{2018Guirado} while no detection at 1.4 GHz. The inferred spectral index of $\alpha$~=~-1.1~$\pm$~0.3 between 8 GHz and 12 GHz is indicative of non-thermal, optically thin, synchrotron, or gyrosynchrotron radiation. Were the 1.4 GHz non-detection due to self-absorption, the magnetic field present in the M7.5 binary would be of 1.2-2.2 kG with a turnover frequency located between 5.0-8.5 GHz. No radio emission was found at the expected position of the L7 object, with a 3$\sigma$ upper limit of 9 $\mu$Jy at 10 GHz.

In this paper we present \textit{Karl G. Jansky} Very Large Array (VLA), European very-long-baseline interferometry (VLBI) Network (EVN), enhanced Multi Element Remotely Linked Interferometer Network (eMerlin), NOrthern Extended Millimeter Array (NOEMA), and Atacama Large Millimetre Array (ALMA) observations of the binary VHS 1256$-$1257AB. Additionally, we re-analysed VLA public data of this system (program 18A-430).
The paper is organized as follows: Section~\ref{sect:observations} describes the observations, Section~\ref{sect:data_red_and_analysis} discusses the data reduction and analysis, Section~\ref{sect:results} presents the results extracted from the observations, Section~\ref{sect:discussion} provides a discussion regarding the various physical constraints that the observations imply, and finally Section~\ref{sect:conclusion} sums up our conclusions.
The analysis and results of the simultaneously observed L7 companion, VHS 1256$-$1257b, will be presented separately (Zakhozhay et al. in prep).


\section{Observations}\label{sect:observations}

We observed the VHS 1256$-$1257 system using the EVN at 5 GHz (4.9350-5.0465 GHz) in phase-referencing mode, with the source J1254-1317 as a phase calibrator. The sequence calibrator-target lasted 4.5 minutes (3.2 minutes on source and 1.3 minutes on the calibrator). The observations were performed in 2018 October in two consecutive days (see Table~\ref{table:obs}).
Both right and left circular polarisations were recorded using eight 16 MHz bandwidth sub-bands per polarisation.

VLA observations were carried in C configuration at 33 GHz with a bandwidth ranging from 29.104 GHz to 36.896 GHz, and using the phase calibrator J1305-1033. The sequence calibrator-target lasted 3.5 minutes (2.5 minutes on source and 1 minute on the calibrator). We recorded right and left circular polarisations with 62 spectral windows of 128 MHz bandwidth each.
We also analysed the VLA public data 18A-430 centered at 6 GHz with bandwidth 3.976-7.896 GHz. This observation was performed using the VLA in A configuration using J1305-1033 as a phase calibrator. Circular polarisations were recorded with 32 spectral windows with 128 MHz bandwidth each. 

The seven-antenna interferometer array eMerlin observed VHS 1256$-$1257 at 5 GHz (4.81-5.33 GHz) in phase-referencing mode, with the source J1305-1033 as a phase calibrator. Observations were performed in three consecutive days in 2017 October. The sequence calibrator-target lasted 10 minutes (7 minutes on source and 3 minutes on the calibrator). Both right and left circular polarisations were recorded.

Sub-mm observations were carried using the NOEMA array in compact configuration D at 230 GHz. The NOEMA field of view was centered at the equidistant point between the central pair and L7 companion. The total on-source time was 1.5~h achieving a 1$\sigma$ rms of 51 $\mu$Jy/beam.

Finally, ALMA observations were carried out 
with 43 of the ALMA 12 m antennas in Band 7 
and a total on-source time of 73 min. 
The longest baseline was 313 m, and the shortest baseline was 15 m long. 
The precipitable water vapor (PWV) in the atmosphere above ALMA was between 1.21 mm and 1.26 mm during the observations.
The observations were obtained at 345 GHz with 7.35 GHz of bandwidth for the continuum, together with spectral line observing mode on baseband 1, centered at the rest frequency of the CO 3-2 line (345.796 GHz) covering a bandwidth of 1.875 GHz with a resolution of 0.98 km/s. 

See Table \ref{table:obs} for further details.

\begin{table*}[t]
    \caption{Journal of observations} 
    \label{table:obs} 
    \centering 
    \begin{tabular}{c c c c c c c c} 
    \hline\hline 
    Array & Frequency (GHz) & Observing Date & UT range & Beam size & P.A ($^{\circ}$) & \begin{tabular}{@{}c@{}}1$\sigma$ rms \\ ($\mu$Jy/beam)\end{tabular}\\ 
    \hline
    eMerlin$^{\mathrm{a}}$ & 5.0 & 
    \begin{tabular}{@{}c@{}c@{}}
    21 Oct 2017 \\ 
    22 Oct 2017 \\ 
    23 Oct 2017 
    \end{tabular} 
    & \begin{tabular}{@{}c@{}c@{}}
    06:17-17:39 \\
    06:35-18:57\\
    06:51-18:58
    \end{tabular}
    & $0.20''\times0.05''$ & 14 & 20\vspace{0.1cm}\\
    EVN$^{\mathrm{b}}$ & 5.0 & \begin{tabular}{@{}c@{}}23 Oct 2018 \\ 24 Oct 2018\end{tabular} & \begin{tabular}{@{}c@{}}06:00-14:00 \\ 06:00-14:00\end{tabular} & 2.46 $\times$ 7.82 mas & 69 & 10
    \vspace{0.1cm}\\
    VLA$^{\mathrm{c}}$ & 6.0 & 13 Apr 2018 & 02:19-06:19 & $0.69''\times0.35''$ & -32 & 3\vspace{0.1cm}\\
    VLA & 33.0 & \begin{tabular}{@{}c@{}}17 Nov 2018 \\ 26 Nov 2018\end{tabular} & \begin{tabular}{@{}c@{}}17:43-18:43 \\ 14:28-15:28\end{tabular} & $1.07''\times0.69''$ & 11 & 7\vspace{0.1cm}\\
    NOEMA & 230 & 24$-$25 Mar 2019 & 23:35-01:36 & $2.53''\times1.43''$ & 0 & 51\vspace{0.1cm}\\
    ALMA & 345 & 7 Mar 2019 & 05:22-06:25 & $0.91''\times0.81''$ & -84.4 & 40\\
    \hline 
    \end{tabular}
    \begin{flushleft}
    \footnotesize{
    \textbf{Notes.} 
    $^{\mathrm{a}}$ eMerlin antennas: Lovell, Mark II, Pickmere, Darnhall, Knockin, Defford, and Cambridge.\\
    $^{\mathrm{b}}$ European VLBI Network using the following antennas: Jodrell Bank, Westerbork, Effelsberg, Medicina, Noto, Onsala, Torun, Yebes, Hartebeesthoek, Cambridge, Darnhall, Defford, Kunming, and Pickmere.\\
    $^{\mathrm{c}}$Public data: program 18A-430.} 
    \end{flushleft}
    \end{table*}

\section{Data reduction and analysis}\label{sect:data_red_and_analysis}

EVN data were reduced using the Astronomical Image Processing System (AIPS) of the National Radio Astronomy Observatory (NRAO) following standard routines. 
The phase-referenced channel-averaged images were deconvolved using the \textit{clean} algorithm implemented in the Caltech imaging software DIFMAP \citep{1994Shepherd} with natural weighting on the visibility data. 
We combined the two observing dates into one data set to improve the signal-to-noise ratio, achieving a 1$\sigma$ rms of 10~$\mu$Jy/beam.
No emission was detected at the expected position.

\begin{figure}
    \centering
    \includegraphics[width=\linewidth]{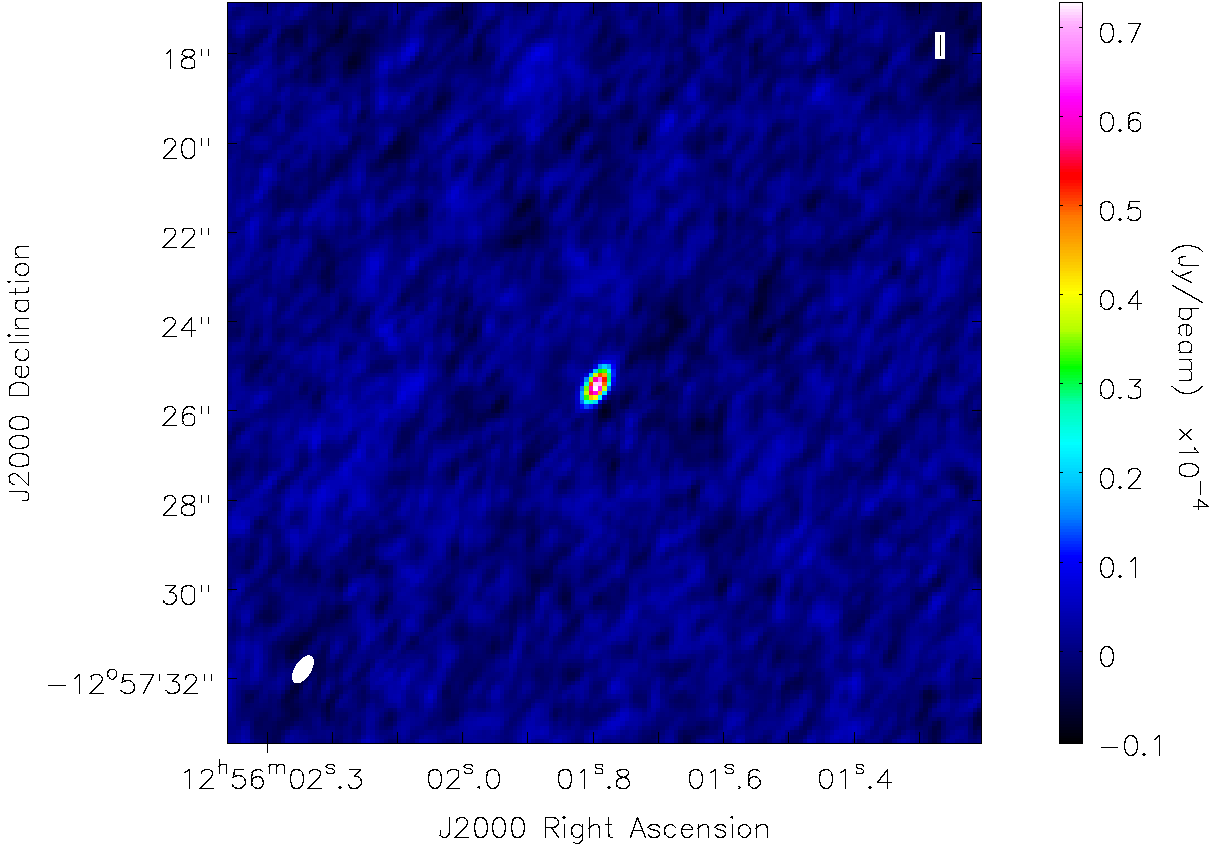}
    \includegraphics[width=\linewidth]{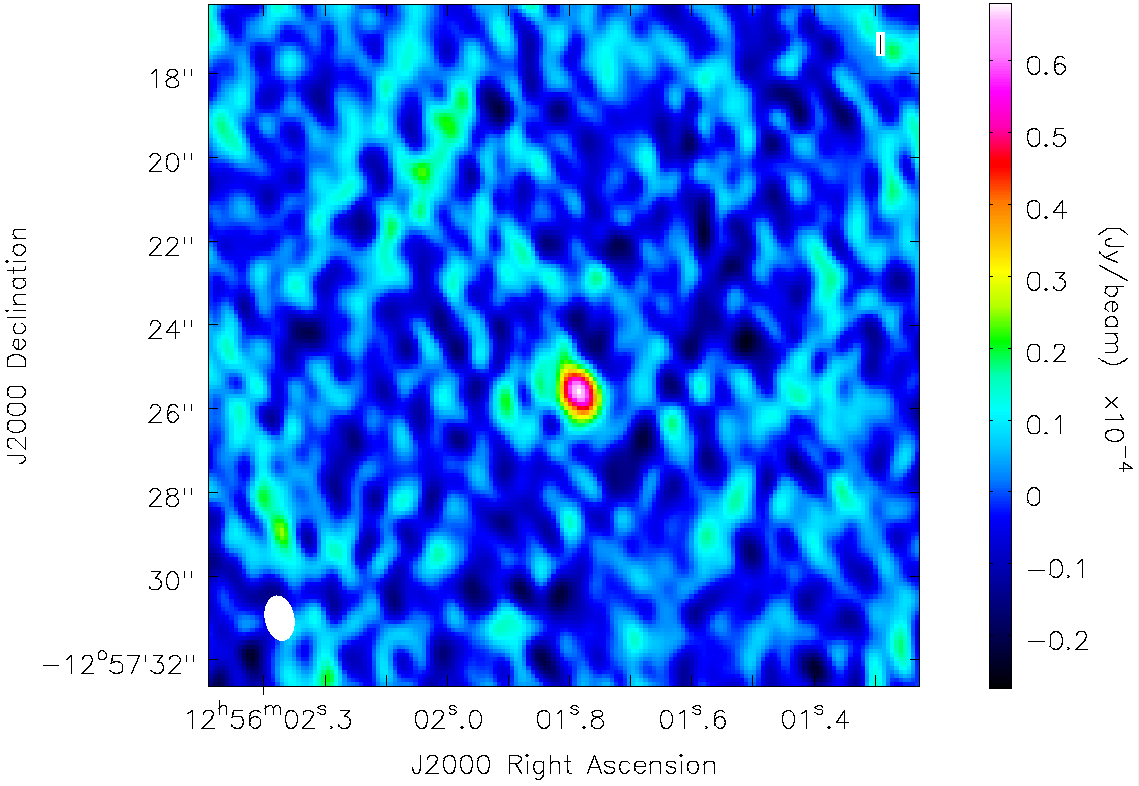}
    \caption{VLA observations of the central binary VHS 1256$-$1257 AB. The size of the radio beam is shown in the lower left corner of each panel.
    \textit{Upper panel}: Stokes I image of the 6 GHz data set.
    \textit{Lower panel}: Same as upper panel but for the 33 GHz data set.}
    \label{fig:mapsCandK}
\end{figure}

VLA data reduction and imaging were carried out using the NRAO CASA\footnote{https://casa.nrao.edu/} software package. The standard procedure of calibration for continuum VLA data was applied. Both 6 GHz and 33 GHz data showed clear detections of the central binary (see Fig.~\ref{fig:mapsCandK}). The achieved 1$\sigma$ rms noise of 3 and 7 $\mu$Jy/beam, respectively, are similar to those of previous VLA observations \citep{2018Guirado}. To further investigate the spectral behaviour of the detected radio emission, we deconvolved  adjacent 360 MHz bandwidth  data  sets  separately for the VLA 6 GHz observations, creating effectively 11 images. Due to the lower signal-to-noise ratio in the 33 GHz detection, only two images could be extracted centered at 31 GHz and 35 GHz, both with a bandwidth of 4 GHz. We repeated this procedure to the published data at 10 GHz \citep{2018Guirado} creating 7 images ranging from 8 to 12 GHz.

We also used CASA for the reduction of eMerlin data, following the standard procedures. To improve the sensitivity, we combined the three consecutive dates into one data set allowing us to reach a 1$\sigma$ rms of 20 $\mu$Jy/beam. No detection was found at the expected position.

NOEMA data calibration was performed with the GILDAS–CLIC software (sep-2019 version)\footnote{http://www.iram.fr/IRAMFR/GILDAS/}. Continuum was obtained by averaging line–free channels over the 7.744~MHz width (USB) centered at 230.0 GHz. No emission was found at the expected position of the central binary VHS 1256$-$1257AB with a 1$\sigma$ rms of 51 $\mu$Jy/beam.

The calibration of the ALMA data followed the standard ALMA Quality Assurance procedure for Cycle 6 
based on the CASA data analysis package version 5.4.$-$70 \citep{2007McMullin}.
We obtained our final image by combining all spectral channels into one image using the \textit{tclean} task of the same CASA package in mfs mode. VHS 1256$-$1257 data were imaged as a single field with a pixel size of 0.16 arcseconds and natural weighting in order to optimize the point-source sensitivity. No significant sources were found in the final image, resulting in a  rms of 40 $\mu$Jy/beam at the location of VHS~1256$-$1257AB.


To extract the flux density of VHS 1256$-$1257AB on each map where the source is detected, we used 
the CASA task \textit{imfit} to fit an elliptical Gaussian with the size of the synthesized beam and centered at the peak intensity. 
We also searched for short-term variability on the detected radio emissions: (i) for 2-day observations (VLA 33 GHz data) we obtained images and flux densities for each day separetely; 
(ii) we analysed the interferometric visibilities, using the AIPS task \textit{DFTPL} which plots the discrete Fourier transform of the complex visibilities for any arbitrary point as a function of time. When necessary, we converted the CASA visibility data set to a UVFITS file using the task \textit{exportuvfits}. To find a balance between signal-to-noise ratio and temporal resolution, we ran \textit{DFTPL} with an interval of 60 seconds.

\section{Results}\label{sect:results}

Radio emission from an unresolved source is detected in both VLA observations. The locations of emission coincide with the expected positions of the binary VHS 1256$-$1256AB, according to the coordinates, proper motion, and parallax given in \citet{2015Gauza}, \citet{2018Guirado}, and \citet{2020Dupuy}. 
The  Stokes I flux density over the whole bandwidth was 73 $\pm$ 4 $\mu$Jy and 83 $\pm$ 13 $\mu$Jy for the 6 GHz and 33 GHz, respectively.  Circularly polarized flux density was not detected at such frequencies. As an upper limit to the fraction of circular polarisation, we computed the value of 3$\times$(rms of Stokes V flux density)/(Stokes I flux density), yielding 0.12 and 0.25 for the 6 GHz and 33 GHz data sets, respectively.

\begin{figure}
    \centering
    \includegraphics[width=\linewidth]{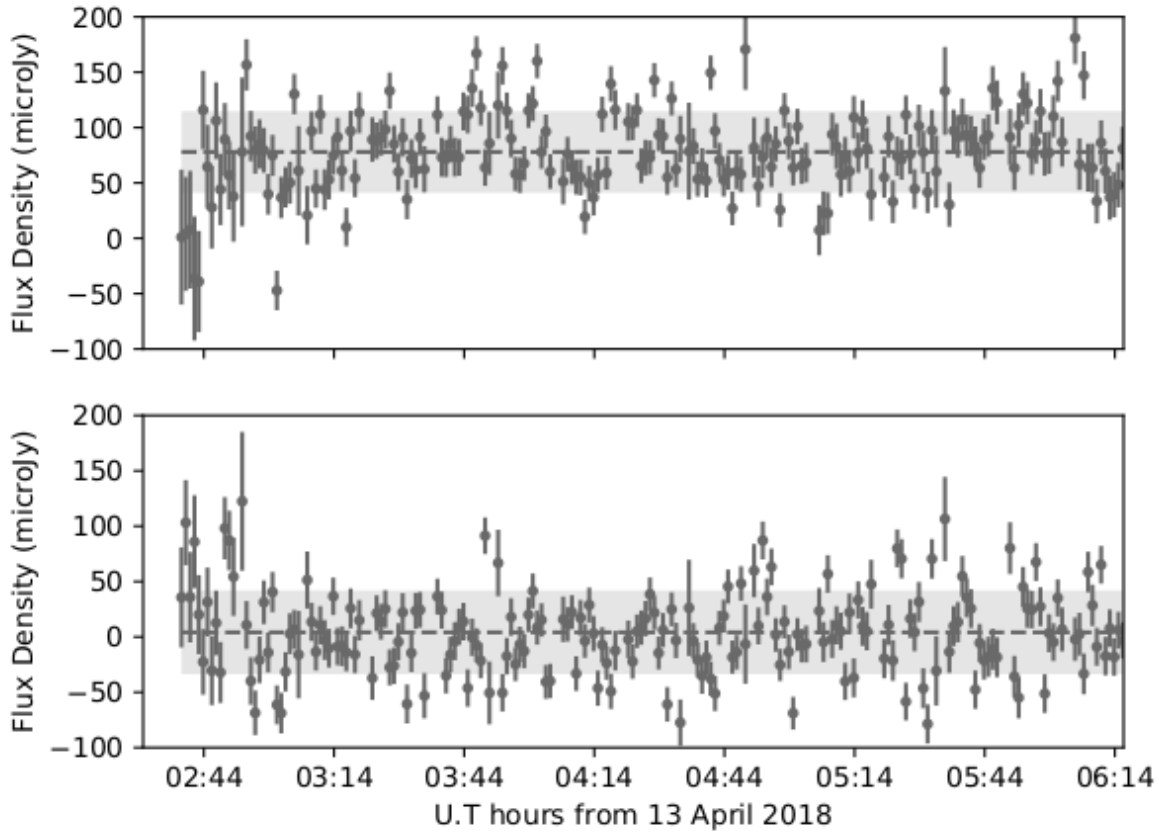}
    \caption{Time  evolution  of  flux  density  in VHS 1256$-$1257 at 6 GHz. 
    \textit{Upper panel}: points show the total flux density (Stokes I) averaged every 60 seconds, with a mean value of 78 $\mu$Jy (plotted as a dashed line) and with a standard deviation of 37 $\mu$Jy (plotted as a grey shadow).
    \textit{Lower panel}: same as upper panel but for Stokes V. The mean and standard deviation values in this case are 4 $\mu$Jy and 38 $\mu$Jy, respectively.}
    \label{fig:Clightcurve}
\end{figure}

\begin{figure*}
    \centering
    \includegraphics[width=\linewidth]{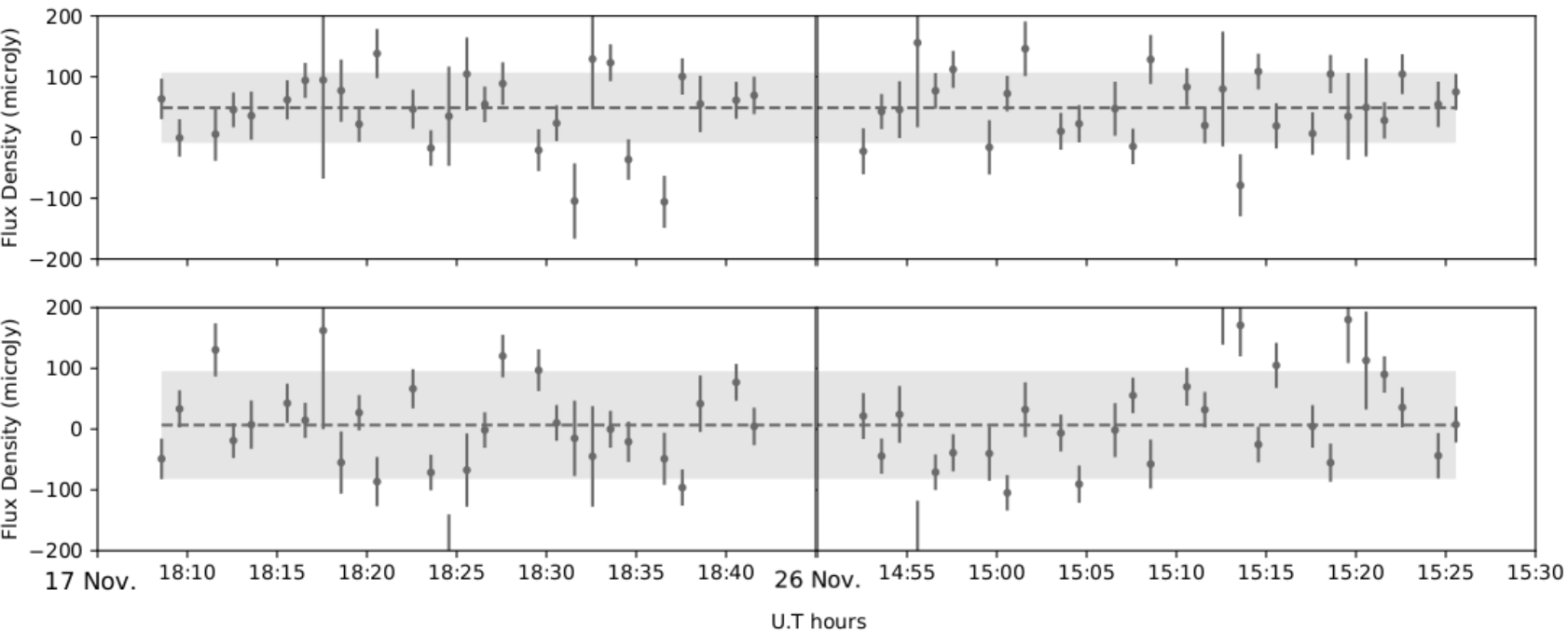}
    \caption{Same as Fig.~\ref{fig:Clightcurve} but for the VLA 33 GHz observations. Mean and standard deviation values are 49 and 58 $\mu$Jy (Stokes I) and 7 and 89 $\mu$Jy (Stokes V).}
    \label{fig:Klightcurve}
\end{figure*}

Figure \ref{fig:Clightcurve} shows the temporal evolution of VHS~1256$-$1257AB flux density (Stokes I and Stokes V) at 6 GHz. The average values are 78 $\mu$Jy and 4 $\mu$Jy with a standard deviations of 37 $\mu$Jy and 38 $\mu$Jy for total and circular flux density, respectively. At the beginning of the observation, a few points show abnormally low total flux density which are likely indicative of some observational effect rather than any physical phenomenon. Throughout the observing time, the total flux density does not present any hints of bursting emission and no significant circular polarisation is seen (as previously anticipated). 
To check for low level periodic signals we computed a generalised Lomb-Scargle periodogram (GLS; see Appendix~\ref{app:periodogram}) and found the maximum power to be a 2$\sigma$ peak with a period of 15.7 $\pm$ 0.3 minutes. We do not consider this peak to be significant, therefore the data are consistent with quiescent emission.
Figure~\ref{fig:Klightcurve} shows a similar plot for the VLA observation at 33 GHz. In this case, no significant temporal variability is detected within each observing day or between them (with a maximum peak in the GLS of 1.3$\sigma$). This indicates that the radio emission from VHS~1256$-$1256AB at this frequency remains stable during, at least, 9 days.

The 6 GHz data clearly matches the spectral behaviour from \citet{2018Guirado} 10 GHz data (see Fig.~\ref{fig:spectrum_radio}). There are two particular images, one with bandwidth from 7.97~GHz to 8.62~GHz (10~GHz published data) and the other from 7.75~GHz to 7.90~GHz (6~GHz new data), that are extremely close in frequency. Despite both observations being almost 3 years apart, the integrated flux densities have remained almost the same: 75~$\pm$~24~$\mu$Jy (15 May 2015) to 73~$\pm$~13~$\mu$Jy (13 April 2018), emphasizing the stability of the radio emission not only on scales of days (as seen in Fig.~\ref{fig:Klightcurve}) but years.
The turnover frequency, i.e. the frequency at which the source transitions from optically thick to optically thin, was estimated to be between 5-8.5~GHz \citep{2018Guirado}. Our new data seems to favour the lower frequency range.

Motivated by previous estimations of the peak emission from the binary at 5.0-8.5~GHz, we obtained VLBI data at 5~GHz. These observations did not detect any compact radio emission at the expected position with a 1$\sigma$ upper limit of 10~$\mu$Jy/beam. We also found no detection of radio emission at 230~GHz (NOEMA) and 345~GHz (ALMA), placing strong upper limits of 51 and 40~$\mu$Jy/beam, respectively. The implications of these non-detections are discussed in the next section.

\section{Discussion}\label{sect:discussion}



\subsection{Constraints from the VLBI no-detections}\label{sect:EVN_nodetection}

The fact that our results show quiescent emission from the central binary that is stable up to almost three years (see Sect.~\ref{sect:results}) renders the time variability scenario unlikely to explain such non-detection. This is because VHS~1256$-$1257AB was observed and detected with the VLA (at a very similar frequency) only $\sim$6 months before the EVN observation, and only a few weeks apart in the case of VLA 33 GHz observations.
Consequently, new scenarios must be considered. The first one that we propose is that of an over-resolved flux component that the EVN and eMerlin are not able to recover due to their higher angular resolution. The second scenario assumes that the detected VLA radio emission comes not from one but from both components of the binary.

Under the assumption that the lack of detection with VLBI is due to an over-resolved component, we can get an estimation for the size of the emitting region by taking into account both VLA 6 GHz and the EVN 5 GHz observations.
From the VLA data set we created an image centered at 5 GHz with a bandwidth of 200 MHz to imitate the EVN observation in terms of central frequency and bandwidth. In this case, we detected VHS~1256$-$1257AB as an unresolved source with a total flux density of 69 $\pm$ 7 $\mu$Jy (rms of 6 $\mu$Jy/beam) and with a synthesized beam of $0.79''\times0.41''$. 
At a distance of 22.2 parsecs, this beam value imposes a maximum size for the emitting region of approximately 13 A.U.
To estimate a lower bound,  we simulated an EVN array with the shortest projected baselines possible and such that the resulting image would have an rms of 14~$\mu$Jy/beam. With this rms, we should be able to detect the 69~$\mu$Jy (5$\sigma$) total flux density recovered from the VLA image, in case of being present. The maximum projected baseline was 2$\times$10$^7$ $\lambda$. With no detection on this image, this implies that the size of the emitting region must be larger than 10~mas (0.22 A.U. or 47 R$_{\odot}$). If we consider a 3$\sigma$ limit then the minimum region size would be $\sim$20~mas (0.44~A.U. or 95~R$_{\odot}$). With the separation between components being 0.1$"$ \citep{2016Stone}, this means that the emitting region size must be between 0.2 and 6 times such separation. We notice that the higher rms of the eMerlin observations prevents us to repeat similar simulations that would have yielded a more stringent low bound.

Additionally, we must consider the scenario where the detected radio emission in our VLA observations comes not from one of the components of the binary but from both. In this case, with a 50~$\mu$Jy/beam 5$\sigma$ rms in our EVN observations and a 69~$\pm$~7~$\mu$Jy total flux density from the VLA equivalent bandwidth image, we estimate the maximum flux ratio between the binary components to be 2.7. The sizes of the emitting regions in this scenario would not need to fit in the values given in the over-resolved scenario. Assuming that both components contribute equally and that they have similar sizes \citep[from 0.12~R$_{\odot}$ to 0.14~R$_{\odot}$, according to the age and mass ranges given in Sect.~\ref{sect:introduction};][]{2000Chabrier}, we calculate a brightness temperature of $4.2-5.7\times10^9$ K for the 6 GHz detection and $1.6-2.1\times10^8$ K for the 33 GHz detection, both values, in principle, consistent  with  synchrotron  or  gyrosynchrotron non-thermal  radio  emission \citep{1985Dulk}. However, the spectral analysis of the data in Fig.~\ref{fig:spectrum_radio} must be taken into account before assigning an emission mechanism, as we will discuss in Sect.~\ref{sect:spectral_analysis}.

Therefore, our VLA detections and VLBI non-detections indicate that either both components of the central binary emit at 5 GHz with a flux ratio $\lesssim$2.7 (in which case no precise estimate of  the  size  of  the  emitting  region can be obtained) or, if the difference is larger i.e. a single component dominates the radio emission, the VLA 6 GHz detection would come from a region with possible sizes raging from $\sim$20~mas ($\sim$730~R$_{*}$) up to $\sim$600~mas ($\sim$21900~R$_{*}$).
These source sizes seem improbable and, consequently, the detected radio emission is likely to be originating in both components with a flux ratio $\lesssim$2.7.

\subsection{Constraints from short-term stability}\label{sect:short_term}

Figure \ref{fig:Clightcurve} demonstrates the lack of short-term variability on this binary during 4 h, indicating that the emitting region is not hidden during such a period of time. 
Recently, the rotation periods for each component have been obtained through the analysis of TESS data \citep[2.0782 $\pm$ 0.0004 h and 2.1342 $\pm$ 0.0003 h; ][]{2021MilesPaez}. 
There is no variability with such periods reproduced in our data (see Sect.~\ref{sect:results}), which indicates that the same emitting region is seen during almost two rotation periods.


The lack of short-term variability at radio wavelengths could be explained if the binary was observed on a pole-on configuration and if the detected radio emission originated in the polar caps of either or both components. In such scenario, no modulation of the radio emission is expected. 
However, \citet{2020Zhou} found that for VHS 1256$-$1257b $\sin{i}\approx$1, that is, the L7 companion is viewed equatorially. 
A similar result was recently found for each component of the central binary \citep{2021MilesPaez}.
Consequently, Fig.~\ref{fig:Clightcurve} should show a rotational modulation as seen in other UCDs. LP 349$-$25 represents an exception to this norm, showing a similar short-term stability during the entire rotation period \citep{2009Osten}. Alternatives scenarios to explain the lack of rotational modulation in LP 349$-$25 are: 
(i) magnetic structures located at the surface of one or both components with a high degree of homogeneity;
(ii) emission coming from both components emitting out of phase so that the detected radio emission remains constant. In the case of VHS~1256$-$1257AB, as in LP 349$-$25, this unlikely scenario would imply that the two components are tidally locked or that information can be communicated on timescales < 60 seconds; 
(iii) a circumbinary radio-emitting structure, which also seems improbable for VHS~1256$-$1257AB given the great separation between components \citep[$\sim$3600 R$_{*}$;][]{2016Stone}. 
We therefore discard this scenario.

An alternative, better-suited scenario capable of explaining the lack of rotational modulation is that where the radio emission is originating at radiation belts around at least one of the components in a similar fashion to 
the magnetospheres of magnetic  chemically  peculiar (MCP) stars.
These main-sequence stars (spectral type A/B) possess a mainly dipolar magnetic field with strengths of $\sim$kG. About 1 in 4 MCP stars emit at radio wavelengths \citep{1994Leone}, and the origin of such radio emission is usually linked to gyrosynchrotron emission from non-thermal electrons moving in a magnetospheric cavity with tens of stellar radii in size \citep[see Fig. 1 of][]{2004Trigilio}.  
In some of these stars, the radio emission is modulated by the stellar rotation as the orientation of the magnetosphere changes with respect to the line of sight as a function of the rotational phase \citep[oblique rotator model;][]{2004Trigilio}. This model has also been applied to the light curve of the UCD 2MASS J13142039+1320011 \citep{2011McLean} and to TVLM 513$-$46546 \citep{2017Leto}.
In the case of VHS~1256$-$1257AB, the total lack of rotational modulation could be explained if the object and belts were seen equatorially (i.e. rotation and magnetic axes perpendicular to our line of sight), which is in agreement with the measurements of \citet{2020Zhou} and \citet{2021MilesPaez}, and would imply no modulation during the entire rotation of the object. 
The size of the radiation belts would not need to be as large as those present in MCP stars, as sizes similar to Jupiter's radiation belts \citep{1981dePater} or slightly larger have been used to describe the field topology of UCD magnetospheres \citep{2017Metodieva}.

Independently of the spatial origin of the radio emission, 
if we assume the typical conditions seen in other UCDs, i.e. radio emission produced by individual bursts originating in a localized region then, in a similar fashion to LP 349$-$25 \citep{2009Osten}, we can constrain the electron density $n_e$ in the case that the energy loss is due to collisions in a high-density environment \citep[Eq. 2.6.20 of][]{2002Benz}. For a temperature >$10^6$ K and a 10 keV particle to have a collisional deflection time >$4$ hours (<60 seconds) requires $n_{e} \lesssim 6 \times 10^{5}$ ($\gtrsim 2 \times 10^{7}$) cm$^{-3}$. If the energy loss is due to radiation losses in a high magnetic field region then the timescales for radiation loss \citep[t$_r$;][]{1985Petrosian} imply magnetic field strengths $B \lesssim 210$ G (t$_r$~>~4 hours) and $B \gtrsim 4$ kG (t$_r$~<~60 seconds) for a 10 keV electron. 
As such, if the origin of the detected radio emission is akin to other UCDs then we constrain the plasma conditions to be either $n_{e} \lesssim 6 \times 10^{5}$ cm$^{-3}$ or $n_{e} \gtrsim 2 \times 10^{7}$ cm$^{-3}$, and $B \lesssim 210$ G or $B \gtrsim 4$ kG.




Therefore, the lack of short-term variability points to either the presence of radiation belts around at least one of the components of VHS 1256$-$1257AB which would be seen equatorially, or to a region on the stellar surface with a high degree of homogeneity. In any of the cases, if the emission region is localized,  the plasma conditions would be constrained by the values given at the end of the previous paragraph. 



\subsection{Constraints from spectral analysis}\label{sect:spectral_analysis}

\begin{figure}
    \centering
    \includegraphics[width=\linewidth]{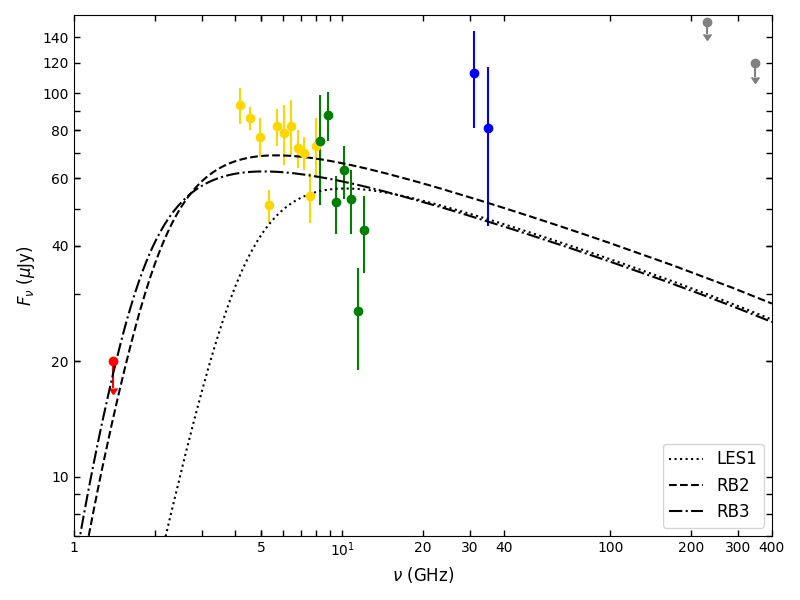}
    \caption{Total flux density of VHS 1256$-$1257AB from VLA observations updated with our new data and with published data \citep{2018Guirado} on a smaller bandwidth sampling (circles). Different colours indicate different bands: 1.4 GHz (red), 6 GHz (yellow), 10 GHz (green), 33 GHz (blue), 230 GHz and 345 GHz (grey). 3$\sigma$ no-detections are plotted as arrows.
    Lines represent different gyrosynchrotron models with $\delta$=1.36 (see Sect.~\ref{sect:flat_index}): dotted line for localized emitting region at the stellar surface with size of 1 R$_*$ (LES1), dashed and dash-dot lines for radiation belts with B$_0$ = 2000 G (RB2) and B$_0$ = 3000 G (RB3), respectively.
    }
    \label{fig:spectrum_radio}
\end{figure}

Quiescent emission of UCDs is successfully interpreted as incoherent gyrosynchrotron emission of energetic electrons in a mild magnetic field \citep[e.g.][]{2001Berger,2009Osten}. At low frequencies (optically thick), the gyrosynchrotron emission spectrum shows a positive slope while at higher frequencies (optically thin) a negative slope.
As previously stated, the VLA detection at 6 GHz (C$-$band) is compatible with the spectrum presented by \citet{2018Guirado}, despite being data taken almost three years apart (see Fig.~\ref{fig:spectrum_radio}). This, together with non-detection at 1.4 GHz (L$-$band), led us to formulate the self-absorption hypothesis in which gyrosynchrotron radiation from  a power-law energy distribution of electrons would explain the observed spectrum between 1$-$12 GHz, with a turnover frequency located around 5 GHz. This mechanism produces a relatively low degree of circular polarisation \citep{1985Dulk} which is also seen in our observations. 
Our 33 GHz (K$_a-$band) detection, however, presents a challenge to this hypothesis as no detectable emission is expected at such frequencies but still a clear component appears in our map (Fig.~\ref{fig:mapsCandK}). To explain the broadband spectrum of VHS~1256$-$1257AB we considered a few scenarios that we shall discuss in turn.


\subsubsection{Single-spectrum hypothesis: flat spectral index}\label{sect:flat_index}


The first scenario consists of an almost flat spectral index, $\alpha$, for the optically thin regime that would encompass the 33 GHz detection. 
This certainly distances from the $\alpha=-1.1$ reported by \citep{2018Guirado} from simultaneous observations between 8 and 12 GHz, but is in an overall better agreement with observations ranging from 4 GHz to 37 GHz, while also explaining the non-detections at NOEMA and ALMA frequencies. Let us discuss the different mechanisms able to produce an almost flat spectral index.

\begin{itemize}
    \item \textit{Thermal bremsstrahlung}. Let $T$ be the temperature of the thermal plasma, $A$ the source area, $L$ the characteristic length scale along the  line  of  sight, $n_e$ the electron density, and VEM the volume emission measure (VEM = $n_e^2 \times A \times L$), the flux density expected for a source located at 22.2 pc due to optically thin thermal bremsstrahlung can be expressed as:


\begin{equation}
\begin{aligned}
S_{\nu}=0.2 \times 10^{-54} \frac{V E M}{\sqrt{T}} \quad J y
\end{aligned}
\end{equation}

where the VEM has units of cm$^{-3}$ and $T$ is in units of K. If the detected radio emission comes from a chromosphere or corona at temperatures 10$^4-$10$^6$ K, then the volume emission measure would range from 5$\times$10$^{52}$ to 5$\times$10$^{53}$ cm$^{-3}$. This range 
is much larger than the X-ray VEM observed in, for example, UV Ceti \citep[2.6 $\times$ 10$^{50}$ cm$^{-3}$;][]{1987Kundu} and other dMe stars \citep[e.g.][]{1990Schmitt,1994White}. 
In those cases where VEM and T are known, optically thin bremsstrahlung radio fluxes from the coronas of UCDs are estimated to be orders of magnitude below the observed fluxes \citep{1981Gary,2000Leto}. 
Unless VHS 1256$-$1257 AB represents a special case, we must conclude that optically thin free-free emission from the corona is probably too weak to produce the measured radio fluxes. We can also dismiss optically thick thermal emission from active regions in the stellar surface or corona akin to $\epsilon$ Eridani \citep{2018Bastian} as the observed brightness temperature in our case is T$_B \gg$ 10$^6$ K (see Sect.~\ref{sect:EVN_nodetection}).

Another possible source of free-free emission is the stellar wind. As an approximation, we assumed an isothermal, spherically symmetric, fully ionized wind originating in one of the components which would have an spectral behaviour $S_{\nu} \propto \nu^{0.6}$ for the optically thick part and $S_{\nu} \propto \nu^{-0.1}$ once the spectrum becomes optically thin \citep{1975Panagia}. 
Using equations 9 and 10 from \citet{2019Rodriguez}, we estimate that a free-free emission from a stellar wind with turnover frequency located at 4 GHz and able to reproduce our 33 GHz detection would imply an electron temperature $T_e \sim 3\times10^5$ K, and a wind mass-loss rate of $\dot{M}_w \sim 2\times10^{-10}$ M$_{\odot}$/yr, under the assumption that the wind velocity is approximately equal to the velocity escape $\sim$520 km/s.
The estimated mass-loss rate is $\sim$1000 greater than 
the predictions from models of winds from cool stars driven by Alfv\'en waves and turbulence \citep{2011Cranmer}, and even greater than the mass-loss rate that would explain the observed spectrum in the much bigger star $\epsilon$ Eridani \citep{2019Rodriguez,2020Suresh}. For this reason, we conclude that the flat spectrum in VHS~1256$-$1257AB cannot be succesfully reproduced by free-free stellar wind emission.

\vspace{0.4cm}

\item \textit{Non-thermal gyrosynchrotron}. The flatness of the spectrum could be reproduced if the detected emission originates in a number of small active regions where self-absorbed non-thermal gyrosynchrotron emits with a different turnover frequency for each region \citep{1989White}. Then it is possible to obtain an approximately flat spectrum with bumps on it. This, however, seems highly unlikely as turnover frequencies as large as 33 GHz would imply magnetic field strengths of $\sim$15 kG using the relation B(kG)~$\approx$~0.151$\nu$(GHz)$^{1.316}$ \citep{1989White}.

Non-thermal gyrosynchrotron emission from a power-law energy distribution of electrons can also reproduce the required flatness of the spectrum. Fixing the optically thin spectral index to $\alpha =$ 0, one can obtain the spectral index of energetic electrons, $\delta$, using the formula \citep{1985Dulk}:

\begin{equation}
\begin{array}{l}
F_{v}=3.3 \times 10^{-24} \times 10^{-0.52 \delta}(\sin \vartheta)^{-0.43+0.65 \delta} \\
\times\left(\frac{v}{v_{B}}\right)^{1.22-0.90 \delta}\left(n_{\mathrm{e}} d\right) B \frac{s}{R^{2}} \mathrm{ergs} / \mathrm{cm}^{2} \mathrm{~s} \mathrm{~Hz}
\end{array}
\end{equation}

where \(\vartheta\) is the angle between the magnetic field and the observer direction, \(v_{\mathrm{B}}\) is the electron gyrofrequency, \(B\) is the magnetic field, \(n_{\mathrm{e}}\) is the concentration of energetic electrons in the source, \(d\) is the thickness of the source projected on the observer, \(s\) is the visible area of the source, and \(R\) is the distance to the radio emission source. 
This would imply a particularly hard spectrum of non-thermal electrons with $\delta$ $\approx$ 1.36 (where $d N / d E \propto E^{- \delta}$). Using this value, we performed numerical simulations of the expected gyrosynchrotron  emission  in the magnetosphere of VHS~1256$-$1257AB (see Appendix~\ref{sect:gyro_modelling}) in two different spatial configurations: a region at the stellar surface, and radiation belts akin to Jupiter's. 
The first one assumes a localized emitting region at the stellar surface which, to concur with the lack of short term variability, must
always be present during our observations, for example, a polar cap seen equatorially. 
As a rough estimation, we fixed the size of the emitting region to be the radius of one component (0.12~R$_{\odot}$, see Sect.~\ref{sect:introduction}). The best fit was found with B~=~140~$\pm$~50~G and n$_e$~=~2.154~$\pm$~0.003~$\times$~10$^4$~cm$^{-3}$, where the error bars indicate the value differences between each simulation. 
The second configuration assumes gyrosynchrotron radio emission originating at radiation belts (see discussion in Sect.~\ref{sect:short_term} and details in Appendix~\ref{sect:gyro_modelling}). In this case, the best values are B~=~140~$\pm$~50~G, n$_e$~=~680~$\pm$~30~cm$^{-3}$ , and B~=~140~$\pm$~50~G and n$_e$~=~320~$\pm$~30~cm$^{-3}$  for a maximum magnetic field strength at the surface level (B$_0$) of 2000 G and 3000~G, respectively.
These models are shown in Fig.~\ref{fig:spectrum_radio} and indicate that a spatial configuration of radiation belts (with approximately the same size as the stellar radius) produces an overall better fit than a localized region.


Independently from its spatial origin, the values for B and n$_e$ obtained with the simulations are in agreement with the estimations that can be made using solely the detections at the extremes of the frequency range, i.e. 5 GHz and 33 GHz.
Typically, gyrosynchrotron emission arises at frequencies $\nu = s\nu_B$, where  the harmonic number, $s$, takes values from 10$-$100 and $\nu_B$ is the electron gyrofrequency given by $\nu_B = 2.8\times10^6B$, with B expressed in Gauss and $\nu_B$ in Hz. For the 5 GHz and 33 GHz detections, gyrosynchrotron emission implies magnetic field strengths in the radio-emitting  source of 18$-$180 G and 120$-$1200 G, respectively. 
In the presence of plasma, gyrosynchrotron emission at harmonic $s$ becomes suppressed at $\nu_p \gtrsim \nu_B s^{-1/2}$ \citep{1985Dulk}. 
Assuming that the 5 GHz detection is caused by this mechanism with harmonics ranging from 10$-$100, this implies $n_e \lesssim$ 3.1$\times$10$^8$ cm$^{-3}$ and $n_e \lesssim$ 3.1$\times$10$^5$ cm$^{-3}$, respectively. Therefore, if the detected 5 GHz radio emission is produced by the gyrosynchrotron mechanism involving high value harmonics (s$\sim$100) then our detection allow us to dismiss the $n_{e} \gtrsim 2 \times 10^{7}$ cm$^{-3}$ estimation given in Sect.~\ref{sect:short_term}, and place an upper limit for the electron density of  $n_e \lesssim$ 3.1$\times$10$^5$ cm$^{-3}$. As such, both constraints for B and n$_e$ are met in our simulations.
For B $\approx$ 140 G, the detection at 33 GHz indicates that the bulk of the gyrosynchrotron emission is originating at s$\sim$80 which implies $n_e~\lesssim$~2.6~$\times$~10$^7$~cm$^{-3}$.

Previously we argued that despite incorporating data taken almost three years apart from our new observations (green points in Fig.~\ref{fig:spectrum_radio}), the part of the spectrum between 4-12~GHz could be treated as if the data were simultaneous.
We ran the simulations for the gyrosynchrotron scenarios described above but utilizing solely data taken a few month apart, that is, the 6~GHz and 33~GHz data sets. The best simulations for each scenario did not differ from those using 6~GHz, 10~GHz, and 33~GHz observations.
As such, a plausible scenario to explain the broad-band spectrum of VHS~1256$-$1257AB is gyrosynchrotron emission coming from radiation belts around (at least) one of the components.
\end{itemize}

Additional scenarios are best explained considering the spectrum as a superposition of a low frequency part (LF) below 15~GHz, and a high frequency part (HF) above 15~GHz. Let us discuss them separately.

\subsubsection{Composite-spectrum hypothesis: The LF regime}\label{sect:LF_regime}

\begin{figure}
    \centering
    \includegraphics[width=\linewidth]{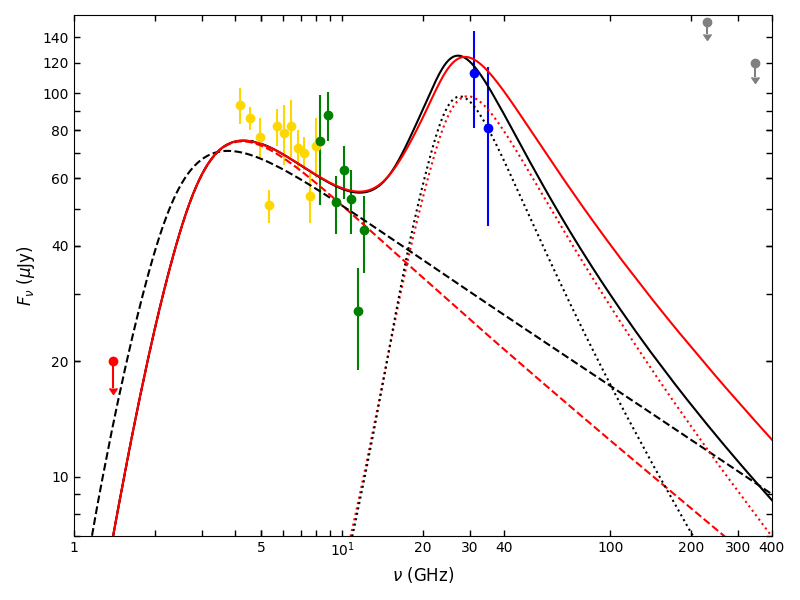}
    \caption{Same as Fig.~\ref{fig:spectrum_radio} but this time lines represent the different gyrosynchrotron models discussed in Sect.~\ref{sect:LF_regime} and Sect.~\ref{sect:HF_regime}: 
    red and black dashed lines for radiation belts at LF with B$_0$~=~2000~G and B$_0$~=~3000~G, respectively;
    red and black dotted line for a localized emission at HF and radiation belts with B$_0$~=~3000~G, respectively;
    red solid line for the combination of radiation belts with B$_0$~=~2000~G at LF and B$_0$~=~3000~G at HF;
    black solid line for the combination of radiation belts with B$_0$~=~2000~G at LF + localized region at HF. 
    }
    \label{fig:spectrum_LF}
\end{figure}

There are two possible mechamisms to explain the LF regime of the spectrum: non-thermal gyrosynchrotron and optically thick thermal gyroresonance.
As previously stated, quiescent emission is usually interpreted as non-thermal gyrosynchrotron emission from a power-law energy distribution of electrons. Limiting the frequencies to the LF part of the spectrum, we repeated the numerical simulations described in Sect.~\ref{sect:flat_index} for the same scenarios, i.e. a 1 stellar radius region at the surface of one component, and radiation belts around at least one component of the binary. However, this time we did not fixed the spectral index of energetic electrons, $\delta$. 
In the case of a localized emitting region, we found no fit better than that presented in Sect.~\ref{sect:flat_index}.
In the case of radiation belts, the best fit corresponds to B~=~140~$\pm$~50~G, n$_e$~=~1.000~$\pm$~0.003~$\times$~10$^4$~cm$^{-3}$, and $\delta$~=~2.1~$\pm$~0.2 for B$_0$~=~2000~G, and B~=~140~$\pm$~50~G, n$_e$~=~1.779~$\pm$~0.003~$\times$~10$^4$~cm$^{-3}$, and $\delta$~=~1.8~$\pm$~0.2 for B$_0$~=~3000~G. These models are shown in Fig.~\ref{fig:spectrum_LF} and indicate that the LF part of the spectrum of VHS 1256$-$1257AB can well be explained with non-thermal gyrosynchrotron emission originating in radiation belts around at least one of the components. 

As stated in Sect.~\ref{sect:flat_index}, the 5~GHz detection implies values of the magnetic field strength in the radio-emitting source B~=~18$-$180~G for gyrosynchrotron emission with harmonics s~=~10$-$100, respectively. 
At a frequency of 12~GHz, this translates to B~=~43$-$430~G. Therefore, the best fitted values for B in any of the scenarios discussed in the simulations are in agreement with these ranges.

An alternative scenario that could explain the LF part of the spectrum is optically thick, thermal gyroresonance radiation, i. e.: non-relativistic electrons spiralling around magnetic field lines. The absorption coefficient for this mechanism is \citep{1985Melrose}:

\begin{equation}
\kappa(\nu)=\frac{(2 \pi)^{3 / 2}}{c} \frac{\nu_{p}^{2}}{\nu} \frac{(s / 2)^{2 s}}{s !} \frac{\sin \theta^{2 s-2}\left(1+\cos ^{2} \theta\right)}{\cos \theta} \beta_{0}^{2 s-3}
\end{equation}

where \(\nu_{p}=\left(n e^{2} /\left(\pi m_{e}\right)\right)^{1 / 2}\) is the plasma frequency, \(s\) the harmonic number, \(\theta\) the angle between the magnetic field and the line of sight, \(\beta_{0}=\left(2 k T /\left(m_{e} c^{2}\right)\right)^{1 / 2}\), and \(\nu_{B}=e B /\left(2 \pi m_{e} c\right)\) is the gyrofrequency.
For an optically thick emission, and assuming a critical optical depth $\tau = \kappa l$ > 2, where \(l=2 \Lambda_{B} \beta_{0} \cos \theta\) is the resonance length, and $\Lambda_B$ the magnetic scale height, \citet{1989Gudel} found that the emission observed in UV Ceti could be reproduced by harmonics $s = 4-6$. \citet{2000Leto} found that the dMe stars V1054 Oph and AU Mic present a similar harmonic range up to where the plasma component is optically thick for gyroresonance emission.
Following these results, and due to the lack of VEM and T measurements, we will take $s = 4$ in our case as a mere illustrative case. 

The observed $\alpha$ $\approx$ -1.1 in the LF part could result from a contraction of the source size with frequency: S$_{\nu} \sim$ r$^2$$\nu^2 \sim \nu^{-1.1}$ from where r $\sim \nu^{-1.55}$. 
Since the frequency of cyclotron emission is proportional to the magnetic field, we have that B $\sim~\nu~\sim$~r$^{-0.65}$. To explain the detections at 4~GHz and 12~GHz, the gyrofrequency equation implies magnetic field strengths between 360~G and 1070~G ($s~=~4$). With $B=B_0~r^{-0.65}$, a photospheric value of $B_0~=~1070$ G implies a source radius of 5.4~R$_*$ for $B=360$~G. This value does not agree with the constraints obtained from Sect. \ref{sect:EVN_nodetection} for emission coming mainly from one of the components of the binary, reinforcing the results that the observed emission is likely coming from both components in the case of optically thick, thermal gyroresonance radiation.
According to equation 5 of \citet{2000Leto}, a source with radius of 5.4~R$_{*}$ would need to have $T~\approx~8.0~\times~10^7$~K to produce the measured emission at 5~GHz (half for equal contribution of both components).
This value is an order of magnitude higher than the plasma temperatures from X-ray data analysis found in the literature for the cool plasma component of dMe stars \citep[e.g.][]{1990Schmitt,1996Giampapa,1999Sciortino} and UCDs \citep[e.g.][]{2009Robrade,2012Stelzer} but similar to the hot plasma component.
This optically thick gyroresonance scenario, however, runs into a problem. \citet{1981Gary} proposed a scaling law for the magnetic field in the outer atmospheres of late type stars where $B \sim r^{-2}$, which was found to be valid for UV Cet, V 1054 Oph, and EV Lac but not for AU Mic where magnetic field strength decreased with radial distance much more rapidly \citep{2000Leto}. Even the slower radial dependence of the magnetic field above active regions in our Sun falls as $B \sim r^{-1.5}$ \citep{1978Dulk}. Therefore, although the current observations cannot rule out optically thick gyroresonance, we warn the reader that, in such case, the presence of a magnetic field with a very low radial dependence needs to be explained.


\subsubsection{Composite-spectrum hypothesis: The HF regime}\label{sect:HF_regime}

Due to the young age of this binary, we considered the possible contribution that a debris disc surrounding VHS~1256$-$1257AB would have to the HF part. In principle, one can find parameters of such disc that would reproduce the density flux recovered in our 33 GHz image. However, this hypothesis was quickly dismissed with the ALMA data. 
We used the BT-Settl photospheric model \citep[][]{2015Baraffe} valid for a cool dwarf with solar metallicity, T$_{\mathrm{eff}}$ = 2600 K, and log g = 5.0 [cm s$^{-2}$], which are the parameters expected for an M7.5 source with an age of a few hundred Myr \citep{2000Chabrier}. This photospheric flux density was then combined with the emission of a putative thin dusty disc with the parameters of the famous AU Mic debris disc \citep{2004Liu}, that is: 0.89 M$_{\mathrm{lunar}}$, grain size of 100 $\mu$m, and temperature of 40~K. We found that the expected flux density at 340~GHz would be ten times larger than our 3$\sigma$ constraint from the ALMA data. Therefore, our observations dismissed the presence of a massive disc similar to that of AU Mic and, consequently, led us to consider alternative scenarios. 


Up to this point, we have assumed that the detected radio emission, if non-thermal gyrosynchrotron, is coming from only one of the components of the binary or from both components but with equal $B$, $n_b$, and $\delta$. We may consider also, the scenario where each component contributes differently not only in terms of integrated flux but in terms of the physical conditions. In this manner, we hypothesize that one component may be the responsible of the radio emission at the LF regime while the other one at the HF regime. To test this hypothesis we ran the gyrosynchrotron simulations described above for the cases of a localized radio emission and a radiation belt with B$_0$~=~3000~G. 
The best fit for a localized emission corresponds to B~=~1760~$\pm$~50~G, n$_e$~=~2.0000~$\pm$~0.0003 $\times$~10$^5$~cm$^{-3}$, $\delta$~=~2.8~$\pm$~0.2. 
For a radiation belt with B$_0$~=~3000~G we found B~=~1050~$\pm$~50~G, n$_e$~=~4.70000~$\pm$~0.00003~$\times$~10$^6$~cm$^{-3}$, and $\delta$~=3.0~$\pm$~0.2 (see Fig.~\ref{fig:spectrum_LF}).
The localized region value for the magnetic field strength does not meet the constraint of Sect.~\ref{sect:short_term}, if the emision is similar to other UCDs. As such, if the emission at 33~GHz is due to non-thermal gyrosynchrotron coming from one of the components, our data favours the spatial origin to be a radiation belt around it.

Can the HF part be explained by thermal gyrosynchrotron? This mechanism is usually expected in highly magnetized stars ($B~\sim$kG) that posses very hot plasma ($T~\sim~10^8$ K).  
Assuming a homogeneous source, this mechanism would produce an optically thick spectral index $\propto$ $\nu^{2}$, which would peak at frequencies $\geq$10~GHz, and a steep optically thin spectral index $\propto$ $\nu^{-8}$ \citep{2002Gudel}. With this spectral behaviour thermal gyrosynchrotron could be responsible for the emission seen at 33~GHz, and given its ultra-steep behaviour, it would be almost imperceptible at NOEMA and ALMA frequencies. 
However, the fact plasma temperatures as high as $10^8$~K are needed and that a very high circular polarisation ($\geq$30\%) is expected from this mechanism \citep{2019Matthews} while no detectable polarisation is seen in VHS 1256$-$1257AB renders thermal gyrosynchrotron very unlikely. 
Could it be then thermal gyroresonance emission akin to that invoked to explain the $U-$shaped spectra of AU Mic \citep{1985Cox}, UV Ceti \citep{1989Gudel}, ER Vul \citep{2003GarciaSanchez}, and other dMe stars \citep{1996Gudel,2000Leto}? 
In these objects, the emission is assumed to come from the X-ray emitting plasma (which is optically thick) and is responsible for the increase in flux seen after 5$-$8~GHz. However, our observations show a decrease in flux not only at 4$-$5~GHz but also at >33 GHz, indicating that the emission at these higher frequencies should come from an optically thin plasma. Is this a reasonable scenario? For a range of objects where both $T$ and VEM are known, \citet{1994White} estimated that thermal gyroresonance is optically thick unless the magnetic scale height, $\Lambda_B~\ll~0.01$~R$_*$. In such case, a clear eclipse signature of the hot X-ray plasma would be expected while none is seen during our observations. Consequently, we conclude that optically thin gyroresonance emission is not a plausible mechanism for the HF part of the spectrum of VHS 1256$-$1257.

One final alternative mechanism to consider is the electron cyclotron maser instability (ECMI) mechanism \citep[as proposed by][to explain the broadband, non-flaring radio emission of UCDs]{2008Hallinan}. This emission would have strong cutoffs at low and high frequencies, justifying the non-detection at 1.4~GHz as well as those at NOEMA and ALMA frequencies. However, our detection at 33 GHz makes this scenario highly unlikely, since it would imply the existence of magnetic field strengths $\geq$~11~kG, which is an order of magnitude higher than that reported in \citet{2018Guirado}, and also much larger than theoretical limits on the surface field strengths of low-mass stars \citep{2014Feiden}. Additionally, this mechanism is expected to produce radio emission with a high degree of circular polarisation, which is not the case for our observations.


Therefore, a distribution of electrons ($n_e~\approx$~10$^{6.5}$~cm$^{-3}$) following a power-law ($\delta~\approx$~3.0) in the presence of a strong magnetic field ($B~\approx$~1 kG) and located at a radiation belt around one of the components of VHS~1256$-$1257AB
would produce non-thermal gyrosynchrotron emission in the HF regime similar to that detected in our observations.


\subsection{Combining all the constraints}\label{sect:all_constraints}


From the discussion of Section \ref{sect:short_term}, and \ref{sect:spectral_analysis} 
%
we conclude that one of the most plausible scenarios that reproduces the broad-brand spectrum of this object is gyrosynchrotron emission from a power-law distribution of electrons where both components of the binary possess different physical conditions ($B$, $n_e$ and $\delta$). When limiting the emission to originate in only one of the components or, equivalently, in both components but with the same $B$, $n_e$ and $\delta$, we found an acceptable fit with emission originating at radiation belts (Fig.~\ref{fig:spectrum_radio}). This hypothesis fits well with the spatial constraint from Sect.~\ref{sect:EVN_nodetection}.
On the other hand, a composite-spectrum hypothesis produces a considerably better fit (Fig.~\ref{fig:spectrum_LF}), particularly  
when we allow each component of the binary to have a different set of free parameters $B$, $n_e$ and $\delta$. This is not unexpected as we are effectively doubling the number of degrees of freedom.
In this scenario, the emission of one of the components would originate in radiation belts peaking at $\sim$4 GHz while emission from the other component would peak at $\sim$30 GHz and would also originate in radiation belts around it.  Although this scenario is in agreement with the lack of short-term variability (Sect.~\ref{sect:short_term}), it is at odds with the spatial constraints obtained from the lack of detection with VLBI (Sect.~\ref{sect:EVN_nodetection}). 
In this configuration, the flux ratio between the binary components at 5~GHz would be much larger than 2.7 and, therefore, the VLA 6~GHz detection would need to come from a region with size of $730~R_* < R < 21900~R_*$;  assuming there is no temporal variability, this size is much larger than the radiation belts diameters which definitely does not favour this scenario.


Another possible scenario capable of concurring with all the constraints discussed above is the combination of optically thick gyroresonance at LF and non-thermal gyrosynchrotron emission at HF. In this case, an explanation for the slow radial dependence of the magnetic field ($B~\sim~r^{-0.65}$) needs to be given for the LF regime. At the HF regime, no strict constraints apply, as a localized emitting region or radiation belts around one or two of the components could be plausible. 

Therefore, if the measured stability in radio emission during almost three years is real, current VLA and VLBI observations of VHS~1256$-$1257AB could be explained by non-thermal gyrosynchrotron radio emission coming from both components and originating at equatorial radiation belts around each UCD. Although a combination of optically thick gyroresonance and non-thermal gyrosynchrotron emission could also be a possible explanation, our analysis renders such scenario less likely.

\section{Conclusions}\label{sect:conclusion}

We have presented new detections of radio emission in the central binary of the substellar triple system VHS 1256$-$1257 centered at 6~GHz and 33~GHz. This emission is not detected with VLBI arrays. 
We have also placed strong upper limits to the radio emission of this binary at 230~GHz and 345~GHz. 
Both detections at 6~GHz and 33~GHz present a Stokes I flux density of 73~$\pm$~4~$\mu$Jy and 83~$\pm$~13~$\mu$Jy, respectively, with no detectable circular polarisation or pulses. 
Let us now gather the constraints that arise from the observations:

\begin{itemize}

    \item The emission at 33~GHz appears stable over a period of 9 days whereas that at 6~GHz is compatible with stability during almost 3 years. 

    \item The lack of detection with VLBI arrays implies that the radio emission produced at 4-8~GHz originates either in both components of the central binary with a flux ratio of $\lesssim$2.7 or in a region with size between $\sim$20~mas and $\sim$600~mas ($\sim$730 and $\sim$21900~R$_{*}$). 
     
    \item The lack of detection with ALMA dismisses the presence of a significant or massive debris disc similar to that of AU Mic.

    \item Both 6~GHz and 33~GHz observation lack any density flux variability during the observations, which has some implications as well. Either the detected flux is emitted in radiation belts around each component of the binary or in a region with a high degree of homogeneity located on the stellar surface.
    
    \item If the emission comes from radiation belts, the rotation and magnetic axes should be perpendicular to our line of sight, that is, they would be seen equatorially.
    
    \item If the emission comes from a localized radio emission (as seen in other UCDs), the plasma conditions must be  $n_{e}~\lesssim~6~\times~10^{5}$~cm$^{-3}$ or $n_{e}~\gtrsim 2~\times~10^{7}$~cm$^{-3}$, and $B~\lesssim~210$~G or $B~\gtrsim~4$ kG.
\end{itemize}

We have discussed various scenarios to explain the spectral behaviour of VHS 1256$-$1257AB. Taking into account the constrains described above, we have narrowed them down to two: 
(i) a combination of optically thick gyroresonance and non-thermal gyrosynchrotron. This scenario seems unlikely as the radial dependence of the magnetic field is uncommonly slow ($B~\sim~r^{-0.65}$) and would need a well-reasoned justification;
(ii) non-thermal gyrosynchrotron mechanism operating at radiation belts around both components of the binary which would be seen equatorially. Plausible conditions for this more likely scenario are: low plasma density (n$_e$~=~300$-$700~cm$^{-3}$), a moderate magnetic field strength  (B~$\approx$~140~G, assuming a $\sim$kG maximum strength on the stellar surface), and a power-law distribution of electrons with $\delta$ fixed at 1.36. 
These values should not be taken as definitive, as the simulations have only been used as a proof of feasibility. New multi-frequency 
observations (at 10$-$30~GHz) of this intriguing system would allow for a more certain determination of both the emission mechanisms involved and the physical parameters, shedding some light into the question of how UCD emit at radio frequencies.
    
\begin{acknowledgements}
We sincerely thank the anonymous referee for his/her very useful and constructive criticisms and suggestions.
This paper is based on observations carried out with the IRAM NOEMA interferometer and the IRAM 30-m telescope. IRAM is supported by INSU/CNRS (France), MPG (Germany), and IGN (Spain).
JBC and JCG were
partially supported by the Spanish MINECO projects  AYA2015-63939-C2-2-P, PGC2018-098915-B-C22 and by the Generalitat Valenciana project GVPROMETEO2020$-$080.
MPT acknowledges financial support from the State Agency for Research of the Spanish MCIU through the "Center of Excellence Severo Ochoa" award to the Instituto de Astrofísica de Andalucía (SEV-2017-0709) and through grants PGC2018-098915-B-C21 and PID2020-117404GB-C21 (MCI/AEI/FEDER, UE).
RA was supported by the Generalitat
Valenciana postdoctoral grant APOSTD/2018/177.
BG acknowledges support from the UK Science and Technology Facilities Council (STFC) via the
Consolidated Grant ST/R000905/1.
MRZO and VJSB acknowledge the financial support from PID2019-109522GB-C51 and PID2019-109522GB-C53, respectively.
\end{acknowledgements}

\bibliographystyle{aa_url} 
\bibliography{references}

\begin{thebibliography}{78}
\expandafter\ifx\csname natexlab\endcsname\relax\def\natexlab#1{#1}\fi

\bibitem[{{Baraffe} {et~al.}(2015){Baraffe}, {Homeier}, {Allard}, \&
  {Chabrier}}]{2015Baraffe}
{Baraffe}, I., {Homeier}, D., {Allard}, F., \& {Chabrier}, G. 2015,
  \href{http://dx.doi.org/10.1051/0004-6361/201425481}{\aap},
  \href{https://ui.adsabs.harvard.edu/abs/2015A&A...577A..42B}{577, A42}

\bibitem[{{Bastian} {et~al.}(2018){Bastian}, {Villadsen}, {Maps}, {Hallinan},
  \& {Beasley}}]{2018Bastian}
{Bastian}, T.~S., {Villadsen}, J., {Maps}, A., {Hallinan}, G., \& {Beasley},
  A.~J. 2018,
  \href{http://dx.doi.org/10.3847/1538-4357/aab3cb}{\apj},
  \href{https://ui.adsabs.harvard.edu/abs/2018ApJ...857..133B}{857, 133}

\bibitem[{{Benz}(2002)}]{2002Benz}
{Benz}, A. 2002, {Plasma Astrophysics, second edition}, Vol. 279

\bibitem[{{Berger}(2002)}]{2002Berger}
{Berger}, E. 2002,
  \href{http://dx.doi.org/10.1086/340301}{\apj},
  \href{https://ui.adsabs.harvard.edu/abs/2002ApJ...572..503B}{572, 503}

\bibitem[{{Berger}(2006)}]{2006Berger}
{Berger}, E. 2006,
  \href{http://dx.doi.org/10.1086/505787}{\apj},
  \href{https://ui.adsabs.harvard.edu/abs/2006ApJ...648..629B}{648, 629}

\bibitem[{{Berger} {et~al.}(2001){Berger}, {Ball}, {Becker}, {Clarke}, {Frail},
  {Fukuda}, {Hoffman}, {Mellon}, {Momjian}, {Murphy}, {Teng}, {Woodruff},
  {Zauderer}, \& {Zavala}}]{2001Berger}
{Berger}, E., {Ball}, S., {Becker}, K.~M., {et~al.} 2001, \nat,
  \href{https://ui.adsabs.harvard.edu/abs/2001Natur.410..338B}{410, 338}

\bibitem[{{Berger} {et~al.}(2009){Berger}, {Rutledge}, {Phan-Bao}, {Basri},
  {Giampapa}, {Gizis}, {Liebert}, {Mart{\'\i}n}, \& {Fleming}}]{2009Berger}
{Berger}, E., {Rutledge}, R.~E., {Phan-Bao}, N., {et~al.} 2009,
  \href{http://dx.doi.org/10.1088/0004-637X/695/1/310}{\apj},
  \href{https://ui.adsabs.harvard.edu/abs/2009ApJ...695..310B}{695, 310}

\bibitem[{{Biller} {et~al.}(2018){Biller}, {Vos}, {Buenzli}, {Allers},
  {Bonnefoy}, {Charnay}, {B{\'e}zard}, {Allard}, {Homeier}, {Bonavita},
  {Brandner}, {Crossfield}, {Dupuy}, {Henning}, {Kopytova}, {Liu},
  {Manjavacas}, \& {Schlieder}}]{2018Biller}
{Biller}, B.~A., {Vos}, J., {Buenzli}, E., {et~al.} 2018,
  \href{http://dx.doi.org/10.3847/1538-3881/aaa5a6}{\aj},
  \href{https://ui.adsabs.harvard.edu/abs/2018AJ....155...95B}{155, 95}

\bibitem[{{Bouy} {et~al.}(2005){Bouy}, {Mart{\'\i}n}, {Brandner}, \&
  {Bouvier}}]{2005Bouy}
{Bouy}, H., {Mart{\'\i}n}, E.~L., {Brandner}, W., \& {Bouvier}, J. 2005,
  \href{http://dx.doi.org/10.1086/426559}{\aj},
  \href{https://ui.adsabs.harvard.edu/abs/2005AJ....129..511B}{129, 511}

\bibitem[{{Browning}(2008)}]{2008Browning}
{Browning}, M.~K. 2008,
  \href{http://dx.doi.org/10.1086/527432}{\apj},
  \href{https://ui.adsabs.harvard.edu/abs/2008ApJ...676.1262B}{676, 1262}

\bibitem[{{Chabrier} {et~al.}(2000){Chabrier}, {Baraffe}, {Allard}, \&
  {Hauschildt}}]{2000Chabrier}
{Chabrier}, G., {Baraffe}, I., {Allard}, F., \& {Hauschildt}, P. 2000,
  \href{http://dx.doi.org/10.1086/309513}{\apj},
  \href{https://ui.adsabs.harvard.edu/abs/2000ApJ...542..464C}{542, 464}

\bibitem[{{Christensen} {et~al.}(2009){Christensen}, {Holzwarth}, \&
  {Reiners}}]{2009Christensen}
{Christensen}, U.~R., {Holzwarth}, V., \& {Reiners}, A. 2009,
  \href{http://dx.doi.org/10.1038/nature07626}{\nat},
  \href{https://ui.adsabs.harvard.edu/abs/2009Natur.457..167C}{457, 167}

\bibitem[{{Cox} \& {Gibson}(1985)}]{1985Cox}
{Cox}, J.~J. \& {Gibson}, D.~M. 1985, {Thermal Emission and Possible Rotational
  Modulation in AU Mic}, ed. R.~M. {Hjellming} \& D.~M. {Gibson}, Vol. 116,
  233--236

\bibitem[{{Cranmer} \& {Saar}(2011)}]{2011Cranmer}
{Cranmer}, S.~R. \& {Saar}, S.~H. 2011,
  \href{http://dx.doi.org/10.1088/0004-637X/741/1/54}{\apj},
  \href{https://ui.adsabs.harvard.edu/abs/2011ApJ...741...54C}{741, 54}

\bibitem[{{de Pater}(1981)}]{1981dePater}
{de Pater}, I. 1981,
  \href{http://dx.doi.org/10.1029/JA086iA05p03423}{\jgr},
  \href{https://ui.adsabs.harvard.edu/abs/1981JGR....86.3423D}{86, 3423}

\bibitem[{{Donati} {et~al.}(2006){Donati}, {Forveille}, {Collier Cameron},
  {Barnes}, {Delfosse}, {Jardine}, \& {Valenti}}]{2006Donati}
{Donati}, J.-F., {Forveille}, T., {Collier Cameron}, A., {et~al.} 2006,
  \href{http://dx.doi.org/10.1126/science.1121102}{Science},
  \href{https://ui.adsabs.harvard.edu/abs/2006Sci...311..633D}{311, 633}

\bibitem[{{Dulk}(1985)}]{1985Dulk}
{Dulk}, G.~A. 1985,
  \href{http://dx.doi.org/10.1146/annurev.aa.23.090185.001125}{\araa},
  \href{https://ui.adsabs.harvard.edu/abs/1985ARA&A..23..169D}{23, 169}

\bibitem[{{Dulk} \& {McLean}(1978)}]{1978Dulk}
{Dulk}, G.~A. \& {McLean}, D.~J. 1978,
  \href{http://dx.doi.org/10.1007/BF00160102}{\solphys},
  \href{https://ui.adsabs.harvard.edu/abs/1978SoPh...57..279D}{57, 279}

\bibitem[{{Dupuy} {et~al.}(2020){Dupuy}, {Liu}, {Magnier}, {Best}, {Baraffe},
  {Chabrier}, {Forveille}, {Metchev}, \& {Tremblin}}]{2020Dupuy}
{Dupuy}, T.~J., {Liu}, M.~C., {Magnier}, E.~A., {et~al.} 2020,
  \href{http://dx.doi.org/10.3847/2515-5172/ab8942}{Research
  Notes of the American Astronomical Society},
  \href{https://ui.adsabs.harvard.edu/abs/2020RNAAS...4...54D}{4, 54}

\bibitem[{{Feiden} \& {Chaboyer}(2014)}]{2014Feiden}
{Feiden}, G.~A. \& {Chaboyer}, B. 2014,
  \href{http://dx.doi.org/10.1088/0004-637X/789/1/53}{\apj},
  \href{https://ui.adsabs.harvard.edu/abs/2014ApJ...789...53F}{789, 53}

\bibitem[{{Fleishman} \& {Kuznetsov}(2010)}]{2010Fleishman}
{Fleishman}, G.~D. \& {Kuznetsov}, A.~A. 2010,
  \href{http://dx.doi.org/10.1088/0004-637X/721/2/1127}{\apj},
  \href{https://ui.adsabs.harvard.edu/abs/2010ApJ...721.1127F}{721, 1127}

\bibitem[{{Gaia Collaboration} {et~al.}(2021){Gaia Collaboration}, {Brown},
  {Vallenari}, {Prusti}, {de Bruijne}, {Babusiaux}, {Biermann}, {Creevey},
  {Evans}, {Eyer}, {Hutton}, {Jansen}, {Jordi}, {Klioner}, {Lammers},
  {Lindegren}, {Luri}, {Mignard}, {Panem}, {Pourbaix}, {Randich}, {Sartoretti},
  {Soubiran}, {Walton}, {Arenou}, {Bailer-Jones}, {Bastian}, {Cropper},
  {Drimmel}, {Katz}, {Lattanzi}, {van Leeuwen}, {Bakker}, {Cacciari},
  {Casta{\~n}eda}, {De Angeli}, {Ducourant}, {Fabricius}, {Fouesneau},
  {Fr{\'e}mat}, {Guerra}, {Guerrier}, {Guiraud}, {Jean-Antoine Piccolo},
  {Masana}, {Messineo}, {Mowlavi}, {Nicolas}, {Nienartowicz}, {Pailler},
  {Panuzzo}, {Riclet}, {Roux}, {Seabroke}, {Sordo}, {Tanga}, {Th{\'e}venin},
  {Gracia-Abril}, {Portell}, {Teyssier}, {Altmann}, {Andrae}, {Bellas-Velidis},
  {Benson}, {Berthier}, {Blomme}, {Brugaletta}, {Burgess}, {Busso}, {Carry},
  {Cellino}, {Cheek}, {Clementini}, {Damerdji}, {Davidson}, {Delchambre},
  {Dell'Oro}, {Fern{\'a}ndez-Hern{\'a}ndez}, {Galluccio}, {Garc{\'\i}a-Lario},
  {Garcia-Reinaldos}, {Gonz{\'a}lez-N{\'u}{\~n}ez}, {Gosset}, {Haigron},
  {Halbwachs}, {Hambly}, {Harrison}, {Hatzidimitriou}, {Heiter},
  {Hern{\'a}ndez}, {Hestroffer}, {Hodgkin}, {Holl}, {Jan{\ss}en}, {Jevardat de
  Fombelle}, {Jordan}, {Krone-Martins}, {Lanzafame}, {L{\"o}ffler}, {Lorca},
  {Manteiga}, {Marchal}, {Marrese}, {Moitinho}, {Mora}, {Muinonen}, {Osborne},
  {Pancino}, {Pauwels}, {Petit}, {Recio-Blanco}, {Richards}, {Riello},
  {Rimoldini}, {Robin}, {Roegiers}, {Rybizki}, {Sarro}, {Siopis}, {Smith},
  {Sozzetti}, {Ulla}, {Utrilla}, {van Leeuwen}, {van Reeven}, {Abbas}, {Abreu
  Aramburu}, {Accart}, {Aerts}, {Aguado}, {Ajaj}, {Altavilla}, {{\'A}lvarez},
  {{\'A}lvarez Cid-Fuentes}, {Alves}, {Anderson}, {Anglada Varela}, {Antoja},
  {Audard}, {Baines}, {Baker}, {Balaguer-N{\'u}{\~n}ez}, {Balbinot}, {Balog},
  {Barache}, {Barbato}, {Barros}, {Barstow}, {Bartolom{\'e}}, {Bassilana},
  {Bauchet}, {Baudesson-Stella}, {Becciani}, {Bellazzini}, {Bernet}, {Bertone},
  {Bianchi}, {Blanco-Cuaresma}, {Boch}, {Bombrun}, {Bossini}, {Bouquillon},
  {Bragaglia}, {Bramante}, {Breedt}, {Bressan}, {Brouillet}, {Bucciarelli},
  {Burlacu}, {Busonero}, {Butkevich}, {Buzzi}, {Caffau}, {Cancelliere},
  {C{\'a}novas}, {Cantat-Gaudin}, {Carballo}, {Carlucci}, {Carnerero},
  {Carrasco}, {Casamiquela}, {Castellani}, {Castro-Ginard}, {Castro Sampol},
  {Chaoul}, {Charlot}, {Chemin}, {Chiavassa}, {Cioni}, {Comoretto}, {Cooper},
  {Cornez}, {Cowell}, {Crifo}, {Crosta}, {Crowley}, {Dafonte}, {Dapergolas},
  {David}, {David}, {de Laverny}, {De Luise}, {De March}, {De Ridder}, {de
  Souza}, {de Teodoro}, {de Torres}, {del Peloso}, {del Pozo}, {Delbo},
  {Delgado}, {Delgado}, {Delisle}, {Di Matteo}, {Diakite}, {Diener},
  {Distefano}, {Dolding}, {Eappachen}, {Edvardsson}, {Enke}, {Esquej}, {Fabre},
  {Fabrizio}, {Faigler}, {Fedorets}, {Fernique}, {Fienga}, {Figueras},
  {Fouron}, {Fragkoudi}, {Fraile}, {Franke}, {Gai}, {Garabato},
  {Garcia-Gutierrez}, {Garc{\'\i}a-Torres}, {Garofalo}, {Gavras}, {Gerlach},
  {Geyer}, {Giacobbe}, {Gilmore}, {Girona}, {Giuffrida}, {Gomel}, {Gomez},
  {Gonzalez-Santamaria}, {Gonz{\'a}lez-Vidal}, {Granvik},
  {Guti{\'e}rrez-S{\'a}nchez}, {Guy}, {Hauser}, {Haywood}, {Helmi}, {Hidalgo},
  {Hilger}, {H{\l}adczuk}, {Hobbs}, {Holland}, {Huckle}, {Jasniewicz},
  {Jonker}, {Juaristi Campillo}, {Julbe}, {Karbevska}, {Kervella}, {Khanna},
  {Kochoska}, {Kontizas}, {Kordopatis}, {Korn}, {Kostrzewa-Rutkowska},
  {Kruszy{\'n}ska}, {Lambert}, {Lanza}, {Lasne}, {Le Campion}, {Le Fustec},
  {Lebreton}, {Lebzelter}, {Leccia}, {Leclerc}, {Lecoeur-Taibi}, {Liao},
  {Licata}, {Lindstr{\o}m}, {Lister}, {Livanou}, {Lobel}, {Madrero Pardo},
  {Managau}, {Mann}, {Marchant}, {Marconi}, {Marcos Santos}, {Marinoni},
  {Marocco}, {Marshall}, {Martin Polo}, {Mart{\'\i}n-Fleitas}, {Masip},
  {Massari}, {Mastrobuono-Battisti}, {Mazeh}, {McMillan}, {Messina},
  {Michalik}, {Millar}, {Mints}, {Molina}, {Molinaro}, {Moln{\'a}r},
  {Montegriffo}, {Mor}, {Morbidelli}, {Morel}, {Morris}, {Mulone}, {Munoz},
  {Muraveva}, {Murphy}, {Musella}, {Noval}, {Ord{\'e}novic}, {Orr{\`u}},
  {Osinde}, {Pagani}, {Pagano}, {Palaversa}, {Palicio}, {Panahi}, {Pawlak},
  {Pe{\~n}alosa Esteller}, {Penttil{\"a}}, {Piersimoni}, {Pineau}, {Plachy},
  {Plum}, {Poggio}, {Poretti}, {Poujoulet}, {Pr{\v{s}}a}, {Pulone}, {Racero},
  {Ragaini}, {Rainer}, {Raiteri}, {Rambaux}, {Ramos}, {Ramos-Lerate}, {Re
  Fiorentin}, {Regibo}, {Reyl{\'e}}, {Ripepi}, {Riva}, {Rixon}, {Robichon},
  {Robin}, {Roelens}, {Rohrbasser}, {Romero-G{\'o}mez}, {Rowell}, {Royer},
  {Rybicki}, {Sadowski}, {Sagrist{\`a} Sell{\'e}s}, {Sahlmann}, {Salgado},
  {Salguero}, {Samaras}, {Sanchez Gimenez}, {Sanna}, {Santove{\~n}a},
  {Sarasso}, {Schultheis}, {Sciacca}, {Segol}, {Segovia}, {S{\'e}gransan},
  {Semeux}, {Shahaf}, {Siddiqui}, {Siebert}, {Siltala}, {Slezak}, {Smart},
  {Solano}, {Solitro}, {Souami}, {Souchay}, {Spagna}, {Spoto}, {Steele},
  {Steidelm{\"u}ller}, {Stephenson}, {S{\"u}veges}, {Szabados}, {Szegedi-Elek},
  {Taris}, {Tauran}, {Taylor}, {Teixeira}, {Thuillot}, {Tonello}, {Torra},
  {Torra}, {Turon}, {Unger}, {Vaillant}, {van Dillen}, {Vanel}, {Vecchiato},
  {Viala}, {Vicente}, {Voutsinas}, {Weiler}, {Wevers}, {Wyrzykowski}, {Yoldas},
  {Yvard}, {Zhao}, {Zorec}, {Zucker}, {Zurbach}, \& {Zwitter}}]{2021Gaia}
{Gaia Collaboration}, {Brown}, A.~G.~A., {Vallenari}, A., {et~al.} 2021,
  \href{http://dx.doi.org/10.1051/0004-6361/202039657}{\aap},
  \href{https://ui.adsabs.harvard.edu/abs/2021A&A...649A...1G}{649, A1}

\bibitem[{{Garc{\'\i}a-S{\'a}nchez} {et~al.}(2003){Garc{\'\i}a-S{\'a}nchez},
  {Paredes}, \& {Rib{\'o}}}]{2003GarciaSanchez}
{Garc{\'\i}a-S{\'a}nchez}, J., {Paredes}, J.~M., \& {Rib{\'o}}, M. 2003,
  \href{http://dx.doi.org/10.1051/0004-6361:20030361}{\aap},
  \href{https://ui.adsabs.harvard.edu/abs/2003A&A...403..613G}{403, 613}

\bibitem[{{Gary} \& {Linsky}(1981)}]{1981Gary}
{Gary}, D.~E. \& {Linsky}, J.~L. 1981,
  \href{http://dx.doi.org/10.1086/159373}{\apj},
  \href{https://ui.adsabs.harvard.edu/abs/1981ApJ...250..284G}{250, 284}

\bibitem[{{Gastine} {et~al.}(2013){Gastine}, {Morin}, {Duarte}, {Reiners},
  {Christensen}, \& {Wicht}}]{2013Gastine}
{Gastine}, T., {Morin}, J., {Duarte}, L., {et~al.} 2013,
  \href{http://dx.doi.org/10.1051/0004-6361/201220317}{\aap},
  \href{https://ui.adsabs.harvard.edu/abs/2013A&A...549L...5G}{549, L5}

\bibitem[{{Gauza} {et~al.}(2015){Gauza}, {B{\'e}jar}, {P{\'e}rez-Garrido},
  {Zapatero Osorio}, {Lodieu}, {Rebolo}, {Pall{\'e}}, \& {Nowak}}]{2015Gauza}
{Gauza}, B., {B{\'e}jar}, V. J.~S., {P{\'e}rez-Garrido}, A., {et~al.} 2015,
  \href{http://dx.doi.org/10.1088/0004-637X/804/2/96}{\apj},
  \href{https://ui.adsabs.harvard.edu/abs/2015ApJ...804...96G}{804, 96}

\bibitem[{{Giampapa} {et~al.}(1996){Giampapa}, {Rosner}, {Kashyap}, {Fleming},
  {Schmitt}, \& {Bookbinder}}]{1996Giampapa}
{Giampapa}, M.~S., {Rosner}, R., {Kashyap}, V., {et~al.} 1996,
  \href{http://dx.doi.org/10.1086/177284}{\apj},
  \href{https://ui.adsabs.harvard.edu/abs/1996ApJ...463..707G}{463, 707}

\bibitem[{{Girard} {et~al.}(2016){Girard}, {Zarka}, {Tasse}, {Hess}, {de
  Pater}, {Santos-Costa}, {Nenon}, {Sicard}, {Bourdarie}, {Anderson},
  {Asgekar}, {Bell}, {van Bemmel}, {Bentum}, {Bernardi}, {Best}, {Bonafede},
  {Breitling}, {Breton}, {Broderick}, {Brouw}, {Br{\"u}ggen}, {Ciardi},
  {Corbel}, {Corstanje}, {de Gasperin}, {de Geus}, {Deller}, {Duscha},
  {Eisl{\"o}ffel}, {Falcke}, {Frieswijk}, {Garrett}, {Grie{\ss}meier}, {Gunst},
  {Hessels}, {Hoeft}, {H{\"o}randel}, {Iacobelli}, {Juette}, {Kondratiev},
  {Kuniyoshi}, {Kuper}, {van Leeuwen}, {Loose}, {Maat}, {Mann}, {Markoff},
  {McFadden}, {McKay-Bukowski}, {Moldon}, {Munk}, {Nelles}, {Norden}, {Orru},
  {Paas}, {Pandey-Pommier}, {Pizzo}, {Polatidis}, {Reich}, {R{\"o}ttgering},
  {Rowlinson}, {Schwarz}, {Smirnov}, {Steinmetz}, {Swinbank}, {Tagger},
  {Thoudam}, {Toribio}, {Vermeulen}, {Vocks}, {van Weeren}, {Wijers}, \&
  {Wucknitz}}]{2016Girard}
{Girard}, J.~N., {Zarka}, P., {Tasse}, C., {et~al.} 2016,
  \href{http://dx.doi.org/10.1051/0004-6361/201527518}{\aap},
  \href{https://ui.adsabs.harvard.edu/abs/2016A&A...587A...3G}{587, A3}

\bibitem[{{Gizis} {et~al.}(2012){Gizis}, {Faherty}, {Liu}, {Castro}, {Shaw},
  {Vrba}, {Harris}, {Aller}, \& {Deacon}}]{2012Gizis}
{Gizis}, J.~E., {Faherty}, J.~K., {Liu}, M.~C., {et~al.} 2012,
  \href{http://dx.doi.org/10.1088/0004-6256/144/4/94}{\aj},
  \href{https://ui.adsabs.harvard.edu/abs/2012AJ....144...94G}{144, 94}

\bibitem[{{G{\"u}del}(2002)}]{2002Gudel}
{G{\"u}del}, M. 2002,
  \href{http://dx.doi.org/10.1146/annurev.astro.40.060401.093806}{\araa},
  \href{https://ui.adsabs.harvard.edu/abs/2002ARA&A..40..217G}{40, 217}

\bibitem[{{Gudel} \& {Benz}(1989)}]{1989Gudel}
{Gudel}, M. \& {Benz}, A.~O. 1989, \aap,
  \href{https://ui.adsabs.harvard.edu/abs/1989A&A...211L...5G}{211, L5}

\bibitem[{{Gudel} \& {Benz}(1996)}]{1996Gudel}
{Gudel}, M. \& {Benz}, A.~O. 1996, in Astronomical Society of the Pacific
  Conference Series, Vol.~93, Radio Emission from the Stars and the Sun, ed.
  A.~R. {Taylor} \& J.~M. {Paredes},
  \href{https://ui.adsabs.harvard.edu/abs/1996ASPC...93..303G}{303}

\bibitem[{{Guirado} {et~al.}(2018){Guirado}, {Azulay}, {Gauza},
  {P{\'e}rez-Torres}, {Rebolo}, {Climent}, \& {Zapatero Osorio}}]{2018Guirado}
{Guirado}, J.~C., {Azulay}, R., {Gauza}, B., {et~al.} 2018,
  \href{http://dx.doi.org/10.1051/0004-6361/201732130}{\aap},
  \href{https://ui.adsabs.harvard.edu/abs/2018A&A...610A..23G}{610, A23}

\bibitem[{{Hallinan} {et~al.}(2008){Hallinan}, {Antonova}, {Doyle}, {Bourke},
  {Lane}, \& {Golden}}]{2008Hallinan}
{Hallinan}, G., {Antonova}, A., {Doyle}, J.~G., {et~al.} 2008,
  \href{http://dx.doi.org/10.1086/590360}{\apj},
  \href{https://ui.adsabs.harvard.edu/abs/2008ApJ...684..644H}{684, 644}

\bibitem[{{Hallinan} {et~al.}(2007){Hallinan}, {Bourke}, {Lane}, {Antonova},
  {Zavala}, {Brisken}, {Boyle}, {Vrba}, {Doyle}, \& {Golden}}]{2007Hallinan}
{Hallinan}, G., {Bourke}, S., {Lane}, C., {et~al.} 2007,
  \href{http://dx.doi.org/10.1086/519790}{\apjl},
  \href{https://ui.adsabs.harvard.edu/abs/2007ApJ...663L..25H}{663, L25}

\bibitem[{{Kao} {et~al.}(2016){Kao}, {Hallinan}, {Pineda}, {Escala},
  {Burgasser}, {Bourke}, \& {Stevenson}}]{2016Kao}
{Kao}, M.~M., {Hallinan}, G., {Pineda}, J.~S., {et~al.} 2016,
  \href{http://dx.doi.org/10.3847/0004-637X/818/1/24}{\apj},
  \href{https://ui.adsabs.harvard.edu/abs/2016ApJ...818...24K}{818, 24}

\bibitem[{{Kirkpatrick} {et~al.}(1997){Kirkpatrick}, {Henry}, \&
  {Irwin}}]{1997Kirkpatrick}
{Kirkpatrick}, J.~D., {Henry}, T.~J., \& {Irwin}, M.~J. 1997,
  \href{http://dx.doi.org/10.1086/118357}{\aj},
  \href{https://ui.adsabs.harvard.edu/abs/1997AJ....113.1421K}{113, 1421}

\bibitem[{{Kundu} {et~al.}(1987){Kundu}, {Jackson}, {White}, \&
  {Melozzi}}]{1987Kundu}
{Kundu}, M.~R., {Jackson}, P.~D., {White}, S.~M., \& {Melozzi}, M. 1987,
  \href{http://dx.doi.org/10.1086/164928}{\apj},
  \href{https://ui.adsabs.harvard.edu/abs/1987ApJ...312..822K}{312, 822}

\bibitem[{{Leone} {et~al.}(1994){Leone}, {Trigilio}, \& {Umana}}]{1994Leone}
{Leone}, F., {Trigilio}, C., \& {Umana}, G. 1994, \aap,
  \href{https://ui.adsabs.harvard.edu/abs/1994A&A...283..908L}{283, 908}

\bibitem[{{Leto} {et~al.}(2000){Leto}, {Pagano}, {Linsky}, {Rodon{\`o}}, \&
  {Umana}}]{2000Leto}
{Leto}, G., {Pagano}, I., {Linsky}, J.~L., {Rodon{\`o}}, M., \& {Umana}, G.
  2000, \aap, \href{https://ui.adsabs.harvard.edu/abs/2000A&A...359.1035L}{359,
  1035}

\bibitem[{{Leto} {et~al.}(2017){Leto}, {Trigilio}, {Buemi}, {Umana},
  {Ingallinera}, \& {Cerrigone}}]{2017Leto}
{Leto}, P., {Trigilio}, C., {Buemi}, C.~S., {et~al.} 2017,
  \href{http://dx.doi.org/10.1093/mnras/stx995}{\mnras},
  \href{https://ui.adsabs.harvard.edu/abs/2017MNRAS.469.1949L}{469, 1949}

\bibitem[{{Lew} {et~al.}(2016){Lew}, {Apai}, {Zhou}, {Schneider}, {Burgasser},
  {Karalidi}, {Yang}, {Marley}, {Cowan}, {Bedin}, {Metchev}, {Radigan}, \&
  {Lowrance}}]{2016Lew}
{Lew}, B. W.~P., {Apai}, D., {Zhou}, Y., {et~al.} 2016,
  \href{http://dx.doi.org/10.3847/2041-8205/829/2/L32}{\apjl},
  \href{https://ui.adsabs.harvard.edu/abs/2016ApJ...829L..32L}{829, L32}

\bibitem[{{Liu}(2004)}]{2004Liu}
{Liu}, M.~C. 2004,
  \href{http://dx.doi.org/10.1126/science.1102929}{Science},
  \href{https://ui.adsabs.harvard.edu/abs/2004Sci...305.1442L}{305, 1442}

\bibitem[{{Liu} {et~al.}(2013){Liu}, {Magnier}, {Deacon}, {Allers}, {Dupuy},
  {Kotson}, {Aller}, {Burgett}, {Chambers}, {Draper}, {Hodapp}, {Jedicke},
  {Kaiser}, {Kudritzki}, {Metcalfe}, {Morgan}, {Price}, {Tonry}, \&
  {Wainscoat}}]{2013Liu}
{Liu}, M.~C., {Magnier}, E.~A., {Deacon}, N.~R., {et~al.} 2013,
  \href{http://dx.doi.org/10.1088/2041-8205/777/2/L20}{\apjl},
  \href{https://ui.adsabs.harvard.edu/abs/2013ApJ...777L..20L}{777, L20}

\bibitem[{{Marois} {et~al.}(2008){Marois}, {Macintosh}, {Barman}, {Zuckerman},
  {Song}, {Patience}, {Lafreni{\`e}re}, \& {Doyon}}]{2008Marois}
{Marois}, C., {Macintosh}, B., {Barman}, T., {et~al.} 2008,
  \href{http://dx.doi.org/10.1126/science.1166585}{Science},
  \href{https://ui.adsabs.harvard.edu/abs/2008Sci...322.1348M}{322, 1348}

\bibitem[{{Marois} {et~al.}(2010){Marois}, {Zuckerman}, {Konopacky},
  {Macintosh}, \& {Barman}}]{2010Marois}
{Marois}, C., {Zuckerman}, B., {Konopacky}, Q.~M., {Macintosh}, B., \&
  {Barman}, T. 2010,
  \href{http://dx.doi.org/10.1038/nature09684}{\nat},
  \href{https://ui.adsabs.harvard.edu/abs/2010Natur.468.1080M}{468, 1080}

\bibitem[{{Matthews}(2019)}]{2019Matthews}
{Matthews}, L.~D. 2019,
  \href{http://dx.doi.org/10.1088/1538-3873/aae856}{\pasp},
  \href{https://ui.adsabs.harvard.edu/abs/2019PASP..131a6001M}{131, 016001}

\bibitem[{{McLean} {et~al.}(2011){McLean}, {Berger}, {Irwin}, {Forbrich}, \&
  {Reiners}}]{2011McLean}
{McLean}, M., {Berger}, E., {Irwin}, J., {Forbrich}, J., \& {Reiners}, A. 2011,
  \href{http://dx.doi.org/10.1088/0004-637X/741/1/27}{\apj},
  \href{https://ui.adsabs.harvard.edu/abs/2011ApJ...741...27M}{741, 27}

\bibitem[{{McMullin} {et~al.}(2007){McMullin}, {Waters}, {Schiebel}, {Young},
  \& {Golap}}]{2007McMullin}
{McMullin}, J.~P., {Waters}, B., {Schiebel}, D., {Young}, W., \& {Golap}, K.
  2007, in Astronomical Society of the Pacific Conference Series, Vol. 376,
  Astronomical Data Analysis Software and Systems XVI, ed. R.~A. {Shaw},
  F.~{Hill}, \& D.~J. {Bell},
  \href{https://ui.adsabs.harvard.edu/abs/2007ASPC..376..127M}{127}

\bibitem[{{Melrose}(1985)}]{1985Melrose}
{Melrose}, D.~B. 1985, {Plasma emission mechanisms.}, ed. D.~J. {McLean} \&
  N.~R. {Labrum}, 177--210

\bibitem[{{Metodieva} {et~al.}(2017){Metodieva}, {Kuznetsov}, {Antonova},
  {Doyle}, {Ramsay}, \& {Wu}}]{2017Metodieva}
{Metodieva}, Y.~T., {Kuznetsov}, A.~A., {Antonova}, A.~E., {et~al.} 2017,
  \href{http://dx.doi.org/10.1093/mnras/stw2597}{\mnras},
  \href{https://ui.adsabs.harvard.edu/abs/2017MNRAS.465.1995M}{465, 1995}

\bibitem[{{Miles-P{\'a}ez}(2021)}]{2021MilesPaez}
{Miles-P{\'a}ez}, P.~A. 2021,
  \href{http://dx.doi.org/10.1051/0004-6361/202141203}{\aap},
  \href{https://ui.adsabs.harvard.edu/abs/2021A&A...651L...7M}{651, L7}

\bibitem[{{Mohanty} {et~al.}(2002){Mohanty}, {Basri}, {Shu}, {Allard}, \&
  {Chabrier}}]{2002Mohanty}
{Mohanty}, S., {Basri}, G., {Shu}, F., {Allard}, F., \& {Chabrier}, G. 2002,
  \href{http://dx.doi.org/10.1086/339911}{\apj},
  \href{https://ui.adsabs.harvard.edu/abs/2002ApJ...571..469M}{571, 469}

\bibitem[{{Morin} {et~al.}(2011){Morin}, {Dormy}, {Schrinner}, \&
  {Donati}}]{2011Morin}
{Morin}, J., {Dormy}, E., {Schrinner}, M., \& {Donati}, J.~F. 2011,
  \href{http://dx.doi.org/10.1111/j.1745-3933.2011.01159.x}{\mnras},
  \href{https://ui.adsabs.harvard.edu/abs/2011MNRAS.418L.133M}{418, L133}

\bibitem[{{Osten} {et~al.}(2009){Osten}, {Phan-Bao}, {Hawley}, {Reid}, \&
  {Ojha}}]{2009Osten}
{Osten}, R.~A., {Phan-Bao}, N., {Hawley}, S.~L., {Reid}, I.~N., \& {Ojha}, R.
  2009,
  \href{http://dx.doi.org/10.1088/0004-637X/700/2/1750}{\apj},
  \href{https://ui.adsabs.harvard.edu/abs/2009ApJ...700.1750O}{700, 1750}

\bibitem[{{Panagia} \& {Felli}(1975)}]{1975Panagia}
{Panagia}, N. \& {Felli}, M. 1975, \aap,
  \href{https://ui.adsabs.harvard.edu/abs/1975A&A....39....1P}{39, 1}

\bibitem[{{Petrosian}(1985)}]{1985Petrosian}
{Petrosian}, V. 1985,
  \href{http://dx.doi.org/10.1086/163765}{\apj},
  \href{https://ui.adsabs.harvard.edu/abs/1985ApJ...299..987P}{299, 987}

\bibitem[{{Radigan} {et~al.}(2013){Radigan}, {Jayawardhana}, {Lafreni{\`e}re},
  {Dupuy}, {Liu}, \& {Scholz}}]{2013Radigan}
{Radigan}, J., {Jayawardhana}, R., {Lafreni{\`e}re}, D., {et~al.} 2013,
  \href{http://dx.doi.org/10.1088/0004-637X/778/1/36}{\apj},
  \href{https://ui.adsabs.harvard.edu/abs/2013ApJ...778...36R}{778, 36}

\bibitem[{{Reiners} \& {Basri}(2010)}]{2010Reinersbasri}
{Reiners}, A. \& {Basri}, G. 2010,
  \href{http://dx.doi.org/10.1088/0004-637X/710/2/924}{\apj},
  \href{https://ui.adsabs.harvard.edu/abs/2010ApJ...710..924R}{710, 924}

\bibitem[{{Rich} {et~al.}(2016){Rich}, {Currie}, {Wisniewski}, {Hashimoto},
  {Brandt}, {Carson}, {Kuzuhara}, \& {Uyama}}]{2016Rich}
{Rich}, E.~A., {Currie}, T., {Wisniewski}, J.~P., {et~al.} 2016,
  \href{http://dx.doi.org/10.3847/0004-637X/830/2/114}{\apj},
  \href{https://ui.adsabs.harvard.edu/abs/2016ApJ...830..114R}{830, 114}

\bibitem[{{Robrade} \& {Schmitt}(2009)}]{2009Robrade}
{Robrade}, J. \& {Schmitt}, J.~H.~M.~M. 2009,
  \href{http://dx.doi.org/10.1051/0004-6361/200811224}{\aap},
  \href{https://ui.adsabs.harvard.edu/abs/2009A&A...496..229R}{496, 229}

\bibitem[{{Rodr{\'\i}guez} {et~al.}(2019){Rodr{\'\i}guez}, {Lizano}, {Loinard},
  {Ch{\'a}vez-Dagostino}, {Bastian}, \& {Beasley}}]{2019Rodriguez}
{Rodr{\'\i}guez}, L.~F., {Lizano}, S., {Loinard}, L., {et~al.} 2019,
  \href{http://dx.doi.org/10.3847/1538-4357/aaf9a6}{\apj},
  \href{https://ui.adsabs.harvard.edu/abs/2019ApJ...871..172R}{871, 172}

\bibitem[{{Route} \& {Wolszczan}(2016)}]{2016Route}
{Route}, M. \& {Wolszczan}, A. 2016,
  \href{http://dx.doi.org/10.3847/2041-8205/821/2/L21}{\apjl},
  \href{https://ui.adsabs.harvard.edu/abs/2016ApJ...821L..21R}{821, L21}

\bibitem[{{Schmitt} {et~al.}(1990){Schmitt}, {Collura}, {Sciortino}, {Vaiana},
  {Harnden}, \& {Rosner}}]{1990Schmitt}
{Schmitt}, J.~H.~M.~M., {Collura}, A., {Sciortino}, S., {et~al.} 1990,
  \href{http://dx.doi.org/10.1086/169525}{\apj},
  \href{https://ui.adsabs.harvard.edu/abs/1990ApJ...365..704S}{365, 704}

\bibitem[{{Sciortino} {et~al.}(1999){Sciortino}, {Maggio}, {Favata}, \&
  {Orlando}}]{1999Sciortino}
{Sciortino}, S., {Maggio}, A., {Favata}, F., \& {Orlando}, S. 1999, \aap,
  \href{https://ui.adsabs.harvard.edu/abs/1999A&A...342..502S}{342, 502}

\bibitem[{{Shepherd} {et~al.}(1994){Shepherd}, {Pearson}, \&
  {Taylor}}]{1994Shepherd}
{Shepherd}, M.~C., {Pearson}, T.~J., \& {Taylor}, G.~B. 1994, in \baas,
  Vol.~26,
  \href{https://ui.adsabs.harvard.edu/abs/1994BAAS...26..987S}{987--989}

\bibitem[{{Shulyak} {et~al.}(2017){Shulyak}, {Reiners}, {Engeln}, {Malo},
  {Yadav}, {Morin}, \& {Kochukhov}}]{2017Shulyak}
{Shulyak}, D., {Reiners}, A., {Engeln}, A., {et~al.} 2017,
  \href{http://dx.doi.org/10.1038/s41550-017-0184}{Nature
  Astronomy}, \href{https://ui.adsabs.harvard.edu/abs/2017NatAs...1E.184S}{1,
  0184}

\bibitem[{{Simitev} \& {Busse}(2009)}]{2009Simitev}
{Simitev}, R.~D. \& {Busse}, F.~H. 2009,
  \href{http://dx.doi.org/10.1209/0295-5075/85/19001}{EPL
  (Europhysics Letters)},
  \href{https://ui.adsabs.harvard.edu/abs/2009EL.....8519001S}{85, 19001}

\bibitem[{{Stelzer} {et~al.}(2012){Stelzer}, {Alcal{\'a}}, {Biazzo},
  {Ercolano}, {Crespo-Chac{\'o}n}, {L{\'o}pez-Santiago},
  {Mart{\'\i}nez-Arn{\'a}iz}, {Schmitt}, {Rigliaco}, {Leone}, \&
  {Cupani}}]{2012Stelzer}
{Stelzer}, B., {Alcal{\'a}}, J., {Biazzo}, K., {et~al.} 2012,
  \href{http://dx.doi.org/10.1051/0004-6361/201118097}{\aap},
  \href{https://ui.adsabs.harvard.edu/abs/2012A&A...537A..94S}{537, A94}

\bibitem[{{Stone} {et~al.}(2016){Stone}, {Skemer}, {Kratter}, {Dupuy}, {Close},
  {Eisner}, {Fortney}, {Hinz}, {Males}, {Morley}, {Morzinski}, \&
  {Ward-Duong}}]{2016Stone}
{Stone}, J.~M., {Skemer}, A.~J., {Kratter}, K.~M., {et~al.} 2016,
  \href{http://dx.doi.org/10.3847/2041-8205/818/1/L12}{\apjl},
  \href{https://ui.adsabs.harvard.edu/abs/2016ApJ...818L..12S}{818, L12}

\bibitem[{{Suresh} {et~al.}(2020){Suresh}, {Chatterjee}, {Cordes}, {Bastian},
  \& {Hallinan}}]{2020Suresh}
{Suresh}, A., {Chatterjee}, S., {Cordes}, J.~M., {Bastian}, T.~S., \&
  {Hallinan}, G. 2020,
  \href{http://dx.doi.org/10.3847/1538-4357/abc004}{\apj},
  \href{https://ui.adsabs.harvard.edu/abs/2020ApJ...904..138S}{904, 138}

\bibitem[{{Trigilio} {et~al.}(2004){Trigilio}, {Leto}, {Umana}, {Leone}, \&
  {Buemi}}]{2004Trigilio}
{Trigilio}, C., {Leto}, P., {Umana}, G., {Leone}, F., \& {Buemi}, C.~S. 2004,
  \href{http://dx.doi.org/10.1051/0004-6361:20040060}{\aap},
  \href{https://ui.adsabs.harvard.edu/abs/2004A&A...418..593T}{418, 593}

\bibitem[{{White} {et~al.}(1989){White}, {Kundu}, \& {Jackson}}]{1989White}
{White}, S.~M., {Kundu}, M.~R., \& {Jackson}, P.~D. 1989, \aap,
  \href{https://ui.adsabs.harvard.edu/abs/1989A&A...225..112W}{225, 112}

\bibitem[{{White} {et~al.}(1994){White}, {Lim}, \& {Kundu}}]{1994White}
{White}, S.~M., {Lim}, J., \& {Kundu}, M.~R. 1994,
  \href{http://dx.doi.org/10.1086/173727}{\apj},
  \href{https://ui.adsabs.harvard.edu/abs/1994ApJ...422..293W}{422, 293}

\bibitem[{{Zakhozhay} {et~al.}(2017){Zakhozhay}, {Zapatero Osorio},
  {B{\'e}jar}, \& {Boehler}}]{2017Zakhozhay}
{Zakhozhay}, O.~V., {Zapatero Osorio}, M.~R., {B{\'e}jar}, V. J.~S., \&
  {Boehler}, Y. 2017,
  \href{http://dx.doi.org/10.1093/mnras/stw2308}{\mnras},
  \href{https://ui.adsabs.harvard.edu/abs/2017MNRAS.464.1108Z}{464, 1108}

\bibitem[{{Zarka}(2000)}]{2000Zarka}
{Zarka}, P. 2000,
  \href{http://dx.doi.org/10.1029/GM119p0167}{Washington DC
  American Geophysical Union Geophysical Monograph Series},
  \href{https://ui.adsabs.harvard.edu/abs/2000GMS...119..167Z}{119, 167}

\bibitem[{{Zechmeister} \& {K{\"u}rster}(2009)}]{2009Zechmeister}
{Zechmeister}, M. \& {K{\"u}rster}, M. 2009,
  \href{http://dx.doi.org/10.1051/0004-6361:200811296}{\aap},
  \href{https://ui.adsabs.harvard.edu/abs/2009A&A...496..577Z}{496, 577}

\bibitem[{{Zhou} {et~al.}(2020){Zhou}, {Bowler}, {Morley}, {Apai}, {Kataria},
  {Bryan}, \& {Benneke}}]{2020Zhou}
{Zhou}, Y., {Bowler}, B.~P., {Morley}, C.~V., {et~al.} 2020,
  \href{http://dx.doi.org/10.3847/1538-3881/ab9e04}{\aj},
  \href{https://ui.adsabs.harvard.edu/abs/2020AJ....160...77Z}{160, 77}

\end{thebibliography}

\newpage
\onecolumn
\begin{appendix}

\section{Gyrosynchrotron modelling}\label{sect:gyro_modelling}

We computed the emission spectra of VHS 1256$-$1257 AB using a fast gyrosynchrotron code \citep{2010Fleishman} in a similar fashion to the procedure described in \citet{2017Metodieva}. Due to our limited data points, we employed the smallest number of parameters possible to describe the source of radio emission. As such, we considered a homogeneous emission source with depth of L and visible area of L$^2$. The energetic electrons are characterized by an isotropic power-law spectrum ($d N / d E \propto E^{- \delta}$) where $\delta$ is known as the spectral index of such distribution. We assumed that this spectrum is valid in the energy range from 10 keV to 100 MeV, which is consistent with previous simulations \citep[see][and references therein]{2017Metodieva}. 
We also assumed the presence of a uniform magnetic field with strength $B$ and a viewing angle relative to the line-of-sight $\theta$.

The fact that we did not detect any Stokes V flux density can be explained by the presence of inhomegeneities in the source. In this case, even if the emission from one region is circularly polarized, it will be compensated by another region with opposite circular polarisation. This is not considered in our models, where we used only Stokes I total flux density and a fixed viewing angle of 80$^{\circ}$. 

Regarding the rest of parameters that define our observations, we let $\delta$ vary from 1.1 to 3.5 (10 passes), B from 10 to 5000~G (40 passes), and n$_e$ from 10$^2$ to 10$^7$~cm$^{-3}$ (20 passes). The two spatial scenarios considered in this work will determine the parameter $L$. 
Firstly, we considered a localized emitting region at the stellar surface we have assumed a size of 0.12~R$_{\odot}$ which corresponds to the estimated radius of one component of the binary. 
Secondly, we considered emission akin to the Jovian decimetric radiation where the gyrosynchrotron mechanism occurs in radiation belts filled with high-energy electrons \citep{2000Zarka}. Following the work of \citet{2017Metodieva}, we approximated these radiation belts as a torus located in the equatorial plane and with volume:

\begin{equation}\label{eq:volume}
\centering
V=2 \pi^{2} R r^{2} \simeq 2 \pi^{2} R\left(R-R_{*}\right)^{2}
\end{equation}

where $R$ and $r$ represent the major and minor radius, respectively. The right hand part of this formula assumes that \(r \simeq R-R_{\mathrm{*}}\), as seen in radio observations of Jupiter \citep[][and references therein]{2016Girard}.

For a dipole-like magnetic field, we can write the average field strength as

\begin{equation}\label{eq:magneticfield}
\centering
B=\frac{B_{0}}{2}\left(\frac{R}{R_{*}}\right)^{-3}
\end{equation}

where $B$ is the magnetic field strength at the minor axis of the torus that represents the radiation belt (at distance $R$ from the UCD centre), and $B_0$ is the maximum surface magnetic field strength. 

As a very rough approximation, we reduce the toroidal source to a homogeneous one so that its volume (Eq. \ref{eq:volume}) $V=L^3$ and its magnetic field strength $B$ is given by Eq. \ref{eq:magneticfield}. For a given magnetic field strength $B$ and assuming $B_0$, the source size can be expressed as follows:

\begin{equation}\label{eq:lsource}
\centering
\frac{L}{R_{*}}= \left(2\pi^2 q \left(q-1\right)^2\right )^{1/3}, \hspace{0.11 cm} \mathrm{where} \hspace{0.1 cm} q=\left(\frac{B_0}{2 B}\right)^{1/3}
\end{equation}

Since UCDs where radio emission is present typically show magnetic fields with the strengths of a few thousand Gauss at the
surface level, we have ran our simulations with $B_0$~=~2000~G and $B_0$~=~3000~G and, to avoid non-physical results, we limited the free parameter B to be lower than B$_0$.


\newpage
\section{GLS periodogram}\label{app:periodogram}

\begin{figure*}[h]
    \centering
    \includegraphics[width=0.9\linewidth]{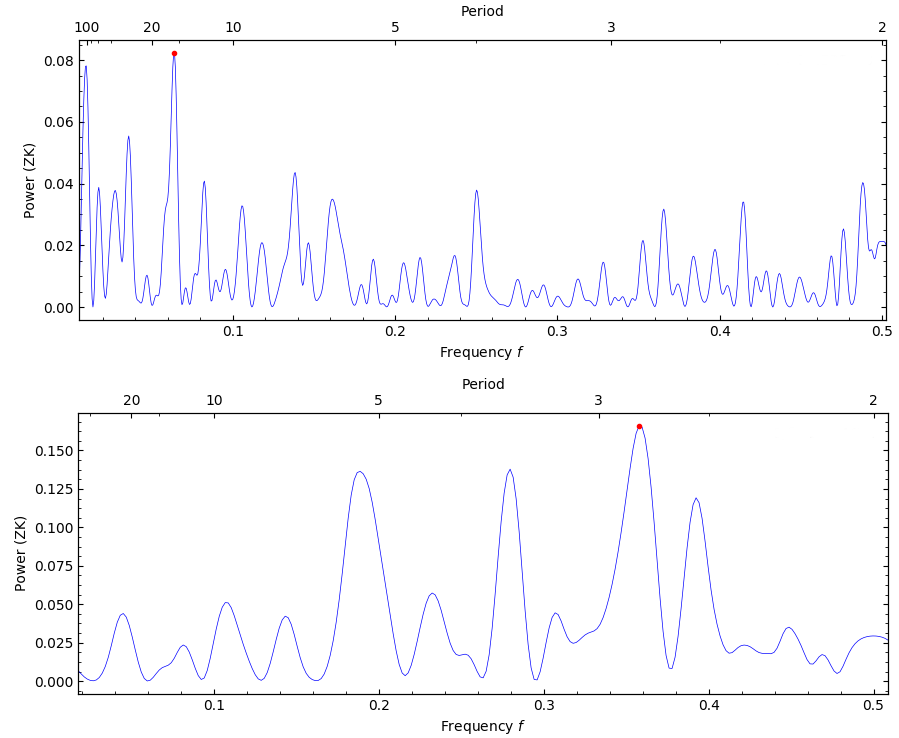}
    \caption{GLS periodogram of 6~GHz Stokes I data set (upper panel) and 33~GHz Stokes I data set (lower panel). Period is given in minutes. The red circles indicate the maximum peak found in the data: 15.7~$\pm$~0.3~minutes at 6~GHz and 2.79~$\pm$~0.05~minutes at 33~GHz.}
    \label{fig:periodogram}
\end{figure*}

To compute the generalised Lomb-Scargle periodogram we used the code provided by \citet[][]{2009Zechmeister}\footnote{https://github.com/mzechmeister/GLS} with the ZK normalisation. We computed such periodogram for the data sets with detections: 6 GHz and 33 GHs Stokes I data sets. The results are shown in Fig.~\ref{fig:periodogram} with maximum peaks of 2$\sigma$ and 1.3$\sigma$ at 15.7~$\pm$~0.3~minutes and 2.79~$\pm$~0.05~minutes, respectively.

\end{appendix}
\end{document}